\documentclass[twocolumn]{tCPH2e}

\usepackage{subfigure}
\usepackage{hyperref}
\theoremstyle{plain}

\theoremstyle{definition}

\theoremstyle{remark}

\begin{document}



\title{Bringing quantum mechanics to life: \\
from Schr\"{o}dinger's cat to Schr\"{o}dinger's microbe}

\author{
\name{Zhang-qi Yin\textsuperscript{a}$^{\ast}$\thanks{$^\ast$Corresponding author. Email: yinzhangqi@tsinghua.edu.cn}
and Tongcang Li\textsuperscript{b,c}$^\dag$\thanks{$^\dag$Corresponding author. Email: tcli@purdue.edu}}
\affil{\textsuperscript{a}Center for Quantum Information, Institute for  Interdisciplinary Information Sciences, Tsinghua University, Beijing 100084, China; \textsuperscript{b}Department of Physics and Astronomy and School of Electrical and Computer Engineering, Purdue University, West Lafayette, IN 47907, USA; \textsuperscript{c}Purdue Quantum Center and Birck Nanotechnology Center, Purdue University, West Lafayette, IN 47907, USA}
\received{August 2016}}

\maketitle

\begin{abstract}
 The question whether quantum mechanics is complete and the nature of the transition between quantum mechanics and classical mechanics have intrigued physicists for decades. There have been many experimental breakthroughs in creating larger and larger quantum superposition and entangled states since Erwin Schr{\"o}dinger proposed his famous thought experiment of putting a cat in a superposition of both alive and dead states in 1935. Remarkably, recent developments in quantum optomechanics and electromechanics may lead to the realization of quantum superposition of living microbes soon. Recent evidence also suggests that quantum coherence may play an important role in several biological processes. In this review, we first give a brief introduction to basic concepts in quantum mechanics and the Schr{\"o}dinger's cat thought experiment. We then review developments in creating quantum superposition and entangled states and the realization of quantum teleportation. Non-trivial quantum effects in photosynthetic light harvesting and avian magnetoreception are also discussed. At last, we review  recent proposals to realize quantum superposition, entanglement and state teleportation of microorganisms, such as viruses and bacteria.

\end{abstract}

\begin{keywords}
Schr\"{o}dinger's cat; quantum superposition; quantum entanglement; quantum teleportation; quantum biology
\end{keywords}

\section{Introduction}
At the beginning of the 20th century, quantum
theory was invented in order to explain  puzzling phenomena related to the black body radiation and atomic emission spectra,
which troubled physicists at that time. After Max Planck introduced the concept of energy quantization to explain the spectrum of the black body radiation in 1901 \cite{Planck}, Albert Einstein  suggested the existence of
light quanta (photon) to explain the photoelectric effect \cite{Einstein05}.
There had been a long debate on whether light was a wave or a group of particles since the age of Isaac Newton.
Einstein's concept of photon provided the  quintessential  example of wave-particle duality, which was later generalized to all matter by  Louis de Broglie in 1924. The de Broglie wavelength of a  particle is $\lambda=h/p$, where $h$ is the Planck constant, and $p$ is the momentum of the particle.
From 1925 to 1927, quantum mechanics was finally formulated into precise mathematical equations,
including the Heisenberg equation and the Schr\"{o}dinger equation. In general, it is not possible to
predict the outcome of a single measurement determinately in quantum mechanics, unless the system is in an eigenstate of the measurement bases.

\begin{figure}[btp]
\begin{center}
\includegraphics[width=10cm]{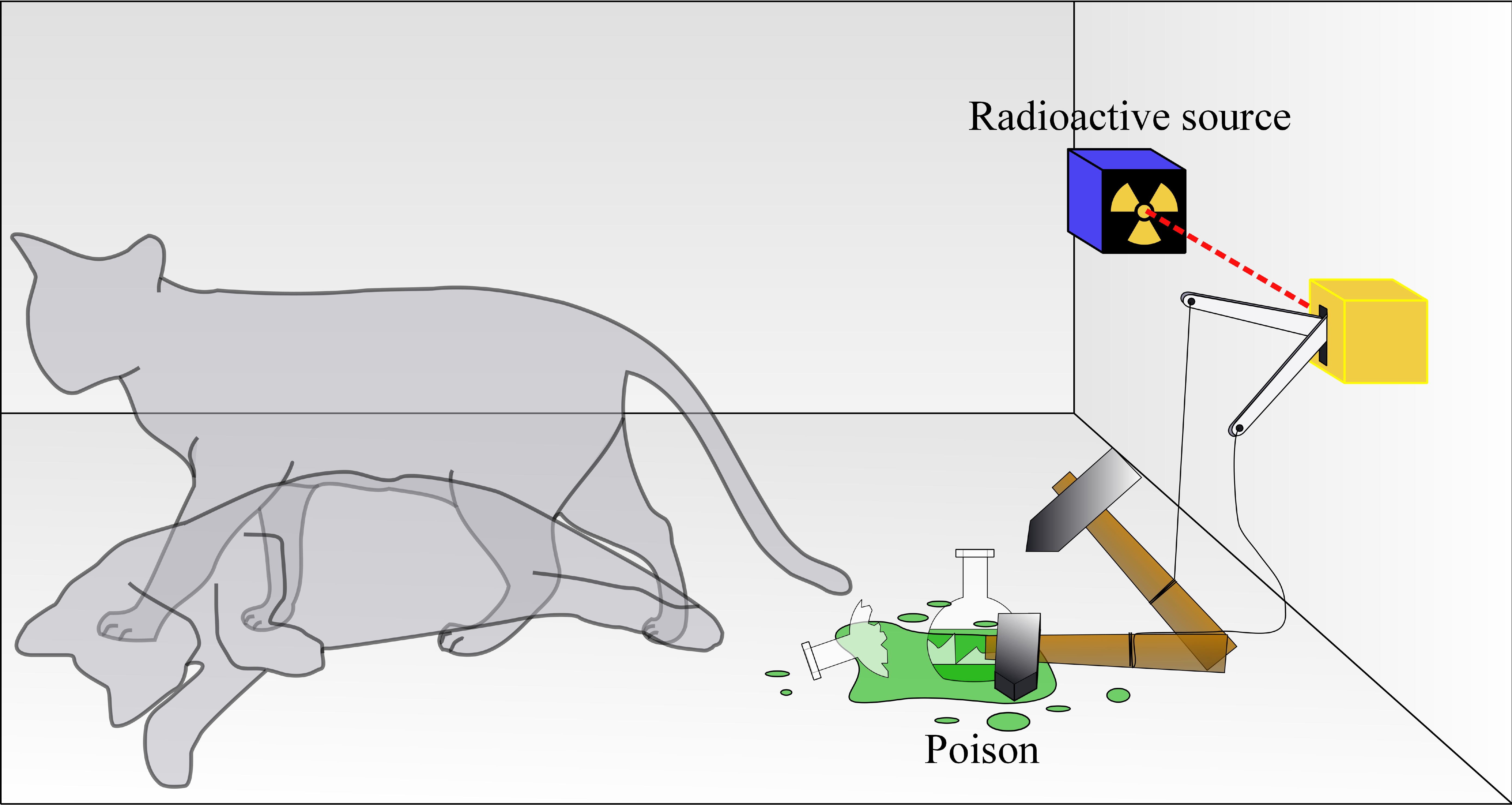}
\caption{Scheme of the Schr\"{o}dinger's cat thought experiment proposed in 1935  \cite{Cat}. A cat, a radioactive source and a bottle of poison are sealed in a box. If an atom in the radioactive source decays, it would trigger a device to release the poison to kill the cat. Thus the living  state of the cat is entangled with the decay state of an atom. At certain time, the atom is in superposition of decay or not decay state, so the cat is simultaneously alive and dead, which violates our common sense. If we open the box, the superposition state of the cat will collapse to either definitely alive or definitely dead in the Copenhagen interpretation. Figure adapted from Wikimedia \cite{CatWiki}.} \label{fig:cat}
\end{center}
\end{figure}

Quantum mechanics has predicted many counterintuitive phenomena which are forbidden in classical mechanics. Since the discovery of quantum mechanics, there were debates on its interpretation. One of the most
famous debates was between A. Einstein and N. Bohr during $1920$s and $1930$s. In a landmark paper
\cite{EPR}, Einstein, Podolsky, and Rosen (EPR) showed that because of the entanglement between distant
particles, either the locality was broken (implied faster-than-light correlation), or the
quantum theory was incomplete. Schr\"{o}dinger exchanged letters with Einstein on the EPR article.
Einstein told Schr\"{o}dinger that the state of an unstable gunpowder could be in the superposition
of both exploded and unexploded states. Schr\"{o}dinger further extended this idea to living systems, such as a cat \cite{Cat}. He proposed to put a living cat in a sealed
chamber, wherein a poison may kill the cat depending on the state of a radioactive atom (Fig. \ref{fig:cat} \cite{CatWiki}). So the macroscopic state of the cat was entangled with the microscopic state of the radioactive atom. After waiting for a certain time, the atom had half chance to be decayed. We can write the wave function of the system as
\begin{equation}\label{eq:schcat}
|\Psi \rangle=\frac{1}{\sqrt{2}}(|{\rm alive}\rangle_{\rm cat} |{\rm undecayed} \rangle_{\rm atom}+|{\rm dead}\rangle_{\rm cat} |{\rm decayed}\rangle_{\rm atom}).
\end{equation}
Following the Copenhagen interpretation of  quantum mechanics, the cat would be both alive and dead
until it was observed. Nowadays, a Schr\"{o}dinger's cat state is generally referred to a quantum superposition state of a macroscopic system that contains multiple degrees of freedom. The states of different degrees of freedom are entangled in a cat state \cite{SchrodingerAtom}.

Schr\"{o}dinger's initial purpose of proposing this thought experiment \cite{Cat} was to illustrate the absurdity of the Copenhagen
interpretation of quantum mechanics, which remains one of the most received interpretations today.
The Schr\"{o}dinger's cat thought experiment stimulated
physicists to propose alternative interpretations of quantum mechanics, such as the many-worlds interpretation initiated
by Hugh Everett \cite{ManyWorld}. Following the many-worlds interpretation, the world is split into two worlds when the Schr\"{o}dinger's
cat is observed. In one world, the cat is alive. But in the other world, the cat is dead. Although the
many-worlds interpretation tries to avoid the conflicts between quantum and classical worlds, it is
almost impossible to be experimentally tested. After proposing the cat thought experiment, Schr\"{o}dinger became interested in explaining biology from the perspective of quantum physics , and wrote a book titled `What is life?' in 1944 \cite{LifeBook}.
This book had a great influence, and stimulated the enthusiasm to search for genetic molecules (such as DNA) in 1950s.

Experimental physicists have tried to realize larger and larger quantum
superposition and entangled states for many years. The superposition of microscopic particles, such as electrons and
atoms, are relatively easy to be generated. There were remarkable progresses in this direction in the past two decades.
In 1996, the Schr\"{o}dinger's cat state was realized with a trapped cold ion \cite{SchrodingerAtom}. A few years later, a matter wave interferometer for
$C_{60}$ molecules was realized \cite{DSC60}. Then, the size of macroscopic quantum systems  increased rapidly with the development of quantum
optomechanics and electromechanics. In 2010,  a mechanical vibration mode of a $30~\mu$m long, 740~nm thick thin film (``quantum drum'')  was cooled
to the quantum regime by  a cryostat, and prepared into quantum superposition states by coupling it with a superconducting qubit \cite{2010Ground}.
Since  the mechanical resonator in the quantum regime is already bigger than many microbes,   quantum superposition of an entire small organism, such as a virus or a bacterium, seems to be feasible as proposed recently \cite{Svirus,Sbacteria}. With the help of a superconducting circuit, the state of a living microbe may also be teleported to another microbe \cite{Sbacteria}. It was proposed that a microbe in a quantum coherent state would represent a new category of cryptobiosis \cite{Bull2015}.
Meanwhile, some organisms seem to be able to harness quantum coherence in biological processes, such as photosynthetic light harvesting and avian magnetoreception, to gain a biological advantage \cite{Lambert2013,Huelga2013}.

In this review, we will briefly summarize experimental and theoretical progresses in realizing Schr\"{o}dinger's cat states, and quantum phenomena in biological systems. In Section \ref{sec:quantum},  basic concepts in
quantum physics and quantum information,
Schr\"{o}dinger's cat thought experiment and relating proposals are introduced. In Section \ref{sec:expsup},
we discuss experimental progresses in quantum superposition and entanglement. We first review how to realize quantum superposition and entanglement in microscopic systems, from single atoms to
complex molecules. Then we discuss recent experiments of generating quantum superposition and entanglement in  optomechanical and electomechanical systems.  In Section \ref{sec:exptel}, we discuss how to realize quantum teleportation with photons, trapped ions, and solid state systems. In Section \ref{sec:Qbio}, we review several biological processes
such as photosynthesis and avian magnetoreception, in which quantum coherence may play an important role. In Section \ref{sec:virus} and \ref{sec:bact}, we discuss two proposals that aim to realize
quantum superposition with living microorganisms. The scheme to teleport the internal state of a microorganism is also reviewed.

\section{Quantum phenomena} \label{sec:quantum}
In this section, we first introduce  basic concepts and terminologies in quantum mechanics and quantum information science. We then introduce the
Schr\"{o}dinger's cat thought experiment and several related thought experiments.

\subsection{Basic concepts in quantum mechanics} \label{sec:concept}
\begin{figure}[btp]
\begin{center}
\includegraphics[width=8cm]{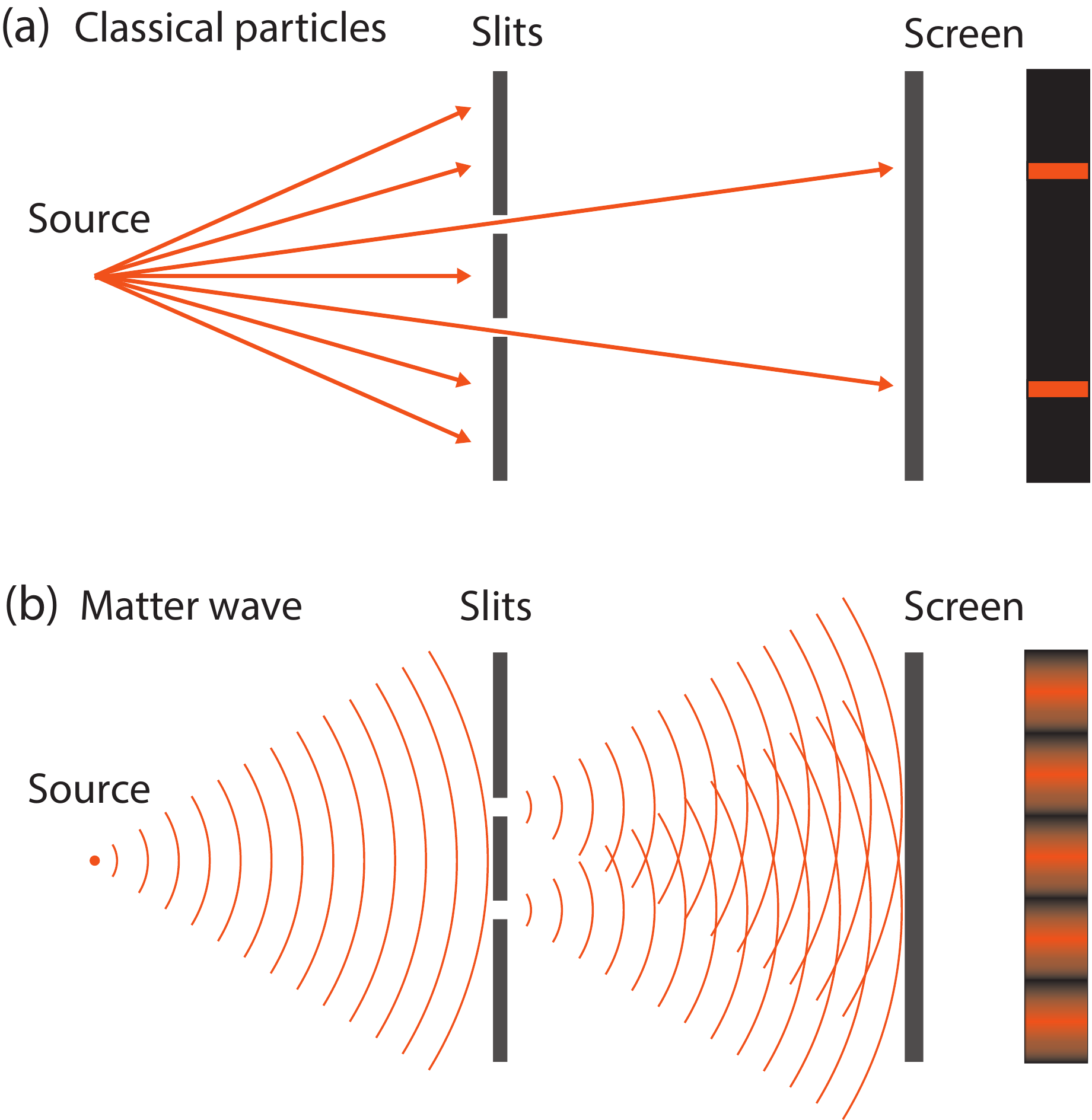}
\caption{Different expectations of the Young's double-slit experiment with atoms in classical mechanics and quantum mechanics. (a) In classical mechanics, each atom has a deterministic trajectory. Only atoms go through one of the slits can reach the screen. We will see two  stripes of atoms on the screen because of the two open slits on the blocking plate. (b) In quantum mechanics, an atom can exhibit wave behaviors. It can go through both the `up' and `down' slits at the same time. We will see an interference pattern of atoms on the screen. } \label{fig:slit}
\end{center}
\end{figure}

An experiment that can drastically distinguish classical mechanics and quantum mechanics would be the Young's double-slit experiment with atoms (or other small particles) (Fig. \ref{fig:slit}). In such an experiment, atoms emitting from a small source with negligible size pass through a thin metal plate with  two open slits. For simplicity, we assume all atoms have the same kinetic energy and mass. In classical mechanics, each atom has a deterministic trajectory. It can only be at one location at a given time. So an atom may be blocked by the plate, or pass through one of the slit. Because we have two open slits on the plate, we will see two stripes of atoms on the screen (Fig. \ref{fig:slit}(a)). In quantum mechanics, however, atoms can have wave behaviors. An atom can pass through both `up' and `down' slits at the same time, which is the state superposition principle.
  We can write the two possible spatial states after an atom
passing through the double slits as $|\text{up}\rangle$ and $|\text{down}\rangle$.
Because of the superposition principle, these two spatial states can form a new superposition state $|\phi\rangle=
c_1 |\text{up}\rangle + c_2 |\text{down}\rangle$. The wavefunction of an atom evolves following the Schr\"{o}dinger equation.
The `up' and `down' parts of the  wavefunction expand, overlap, and finally form an interference pattern on the screen (Fig. \ref{fig:slit}(b)).
Traditionally, the superposition principle was only used for microscopic
particles, such as electrons, atoms and molecules. Recent experiments showed that it can also be applied for macroscopic
systems, such as micro-mechanical resonators \cite{2010Ground}.

 By  applying the superposition principle mathematically, a
 quantum bit (qubit), a  fundamental concept in quantum information science,  can be defined. Just like a classical
 bit has a state either $0$ or $1$, a qubit can be in state $|0\rangle$ or $|1\rangle$. Unlike the
classical bit,  a qubit can also be in the state
 \begin{equation}\label{eq:qubit}
   |\psi \rangle = \alpha |0\rangle + \beta |1\rangle,
 \end{equation}
 which is an arbitrary superposition of $|0\rangle$ and $|1\rangle$. After measurement, the qubit could be either in
$|0\rangle$ state with probability $|\alpha|^2$, or in $|1\rangle$ state with probability $|\beta|^2$. The complex numbers
 $\alpha$ and $\beta$ must fulfill the normalization condition $|\alpha|^2 + |\beta|^2 =1$. Generally speaking, a qubit state
is a unit vector in a two-dimensional vector space.

Based on the superposition principle, entanglement was introduced and discussed by Einstein, Podolsky, and Rosen in 1935 \cite{EPR}. They found that two particles
could be prepared in a special
state that cannot be described by two separated individual particle states, no matter how far the particles were separated. In other words,
these two particles seemly correlate with ``spooky action at a distance'' \cite{Einstein1947}.
The debates on the EPR paradox led John Bell to define an inequality to distinguish a theory with local hidden variables from the quantum mechanics \cite{Bell1964}.
Experimental tests of the  Bell inequality have continued for more than 40 years, in order to close  all known loopholes. In 2015, three experiments,
which were performed with nitrogen-vacancy centers and single photons systems, closed all known loopholes and verified the Bell inequality. They excluded local
hidden variable theories \cite{Hensen2015,Shalm2015,Giustina2015} and proved the faster-than-light correlation in quantum entanglement experimentally.

Before defining an entangled state, we should first introduce the concept of separable states.
Let's focus on  two qubits $A$ and $B$, whose Hilbert spaces are $H_A$ and
$H_B$. For the whole system that includes two qubits, the Hilbert space is $H_A \otimes H_B$.
If two qubis are in pure states $|\psi_A\rangle= \alpha_A |0\rangle_A + \beta_A|1\rangle_A$ and $|\psi_B\rangle= \alpha_B |0\rangle_B +
\beta_B |1\rangle_B$, the system of the composite system is
$|\psi\rangle_{AB}= |\psi_A\rangle \otimes |\psi_B\rangle = (\alpha_A|0\rangle_A + \beta_A
|1\rangle_A) \otimes ( \alpha_B |0\rangle_B + \beta_B |1\rangle_B)$,
which is a separable state. However, the most general states  of the composite system is
\begin{equation}
|\psi\rangle_{AB}= \sum_{i,j=0}^1 c_{i,j} |i\rangle_A \otimes |j\rangle_B,
\end{equation}
which cannot always be represented in the form of product states. For a state that is inseparable, we call it an entangled state.

Let's take a Bell state $\frac{1}{\sqrt{2}} (|0\rangle_A |1\rangle_B + |1\rangle_A |0\rangle_B)$ as an example of entangled states. The reduced
density matrix for either $A$ or $B$ is totally mixed. Let an observer Alice measure the system $A$, and observer Bob measure the system $B$ on the
computational bases $|0\rangle$ and $|1\rangle$. Both Alice and Bob will get random outcomes $0$ or $1$.
If we combine their outcomes together, however, we will find that these outcomes are totally correlated. When Alice gets the outcome $0$, Bob must get
the outcome $1$, and vice versa. Unlike  classical bits, we can also measure the qubits on the superposition basis $|\pm \rangle= (|0\rangle
\pm |1\rangle)/\sqrt{2}$. The outcomes between $A$ and $B$ are also totally correlated, no matter how large the separation between the two qubits.
Therefore, the entangled states can only be described for the whole system, other than the individual subsystems.

As the speed of correction in quantum entanglement is much faster than the light speed, can it be used for faster-than-light communication?
Unfortunately, the answer is no.
The next question is, can we use this correlation as a resource in communication? The answer is yes. In 1993, quantum teleportation was proposed \cite{tel}. In quantum
teleportation, quantum entanglement is used as a resource for transferring an unknown quantum state from one location to another without physically
moving the particles that the state is stored. The  quantum teleportation also requires the assistance of classical communication.
Therefore, the speed of information transmission cannot be faster than the speed of light in quantum teleportation. We will discuss more details of the quantum teleportation protocol in
Section \ref{sec:TelPro}.

\subsection{Schr\"{o}dinger's cat thought experiment}

Here we discuss the Schr\"{o}dinger's cat thought experiment and review  relating thought experiments.
The original version of the Schr\"{o}dinger's cat thought experiment \cite{Cat} is shown in Fig. \ref{fig:cat}. A cat is  in a sealed chamber, along with
a flask of poison which may kill the cat. There is also a tiny bit of radioactive material. In an hour or so, there is $50\%$ possibility that an atom may decay, and
$50\%$ possibility that no decay occurs. A Geiger counter, once it measures a radioactive decay event, would trigger a lethal device that releases the poison and kills
the cat. In other words, the cat's living state is entangled with the atomic state. The wave function of the system is $|\Psi \rangle=\frac{1}{\sqrt{2}}(|{\rm alive}\rangle_{\rm cat} |{\rm undecayed} \rangle_{\rm atom}+|{\rm dead}\rangle_{\rm cat} |{\rm decayed}\rangle_{\rm atom})$.
Following the Copenhagen interpretation,  the alive or dead state of the cat is
 not determined before we open the chamber and observe the cat. When we open the chamber, however, the state of the cat becomes definitely alive or definitely dead.
As an extension to the EPR paradox, the Schr\"{o}dinger's cat thought experiment shows the possibility to entangle  a macroscopic system with a microscopic system.
It also reveals the incompleteness of the measurement theory in the Copenhagen interpretation. It seems that the life of the cat is determined by
the observer.

Since the cat thought experiment was proposed, several interpretations have been proposed to resolve this paradox. A famous one
is the many-worlds interpretation \cite{ManyWorld}.  The many-worlds interpretation treats the whole universe as a wavefunction.
Unlike the Copenhagen interpretation, the many-worlds interpretation does not treat measurement as a non-unitary process, and denies
the wavefunction collapse. When a measurement is performed, all the different outcomes are obtained, and each outcome is in a different world.
 When we open the chamber and observe the Schr\"{o}dinger's cat, the universe will split into two different ones.
The alive and dead cats will be in different universes. Both universes are real in the many-worlds interpretation.

An interesting extension of the Schr\"{o}dinger's cat thought experiment is to add an observer in the chamber to observe the cat's state. In this way,
the quantum mechanics is applied to ourselves, especially our consciousness, and leads to even  stranger conclusions \cite{Wigner,Deutsch86,VV2015}.
The key point is that we can communicate with the observer inside the chamber before we open it. We may ask the observer whether the cat is in a
definite state (can be dead or alive). If the observer answers ``Yes'', we may undo the experiment since the evolution is reversible in quantum mechanics.
We should note that the answer of the observer does not collapse the wavefunction to a definite ``dead'' or ``alive'' state as we still do not know whether the cat is dead or alive from the answer.  If the cat
is dead, we may make it alive again by reversing the quantum process. So the poison is back into the bottle, the atom does not decay, and the observer does not remember seeing a dead cat.
However, for the observer inside the chamber, his observation should make the cat's wave function collapse. His answer is the proof of it. In this way, a contradiction appears.

Doing such experiments with a cat is impossible right now. Therefore, people started to realize quantum superposition using relatively small objects.
Up to now, single atoms, molecules with many atoms, and even micrometer-scale mechanical resonators have been prepared into quantum superposition
states. Some of them  were even generated into entangled states. We will summarize these progresses in Section \ref{sec:expsup}.
Schemes to realize quantum superposition of living microorganisms,
such as viruses and bacteria, have been proposed \cite{Svirus,Sbacteria}. Ref. \cite{Sbacteria} also discussed how to teleport the internal electron spin state of a living
bacterium to the internal electron spin state of another remote bacterium. In Section \ref{sec:virus} and \ref{sec:bact}, we will review these proposals
and related experimental progresses.

\section{Experimental developments in quantum superposition and entanglement} \label{sec:expsup}

In this section, we will first
discuss the experimental developments in superposition and entanglement of atoms  and molecules. Then we will discuss quantum behaviors of larger objects,
such as optomechanical systems.

\subsection{Matter wave interferometry}

In quantum mechanics, Young's double-slit experiment is commonly used to study the wave feature of  matter, which is
also a proof of the quantum superposition of particle spatial states after the double slit. Double-slit experiments have been repeated with larger and larger particles. The first example was the double-slit experiment with single electrons,
which was demonstrated more than $40$ years ago \cite{DSeletron}. In 2002, this experiment was selected as the most beautiful experiment
in physics by ``Physics World" \cite{beautiful}.

\begin{figure}[btp]
\begin{center}
\begin{minipage}{130mm}
\subfigure[A double-slit interferometer for atoms \cite{DSatom}. The atoms can go through two different pathes from the source to the detector. The distance between the source and the detector is 128 cm. ]{
\resizebox*{6.5cm}{!}{\includegraphics{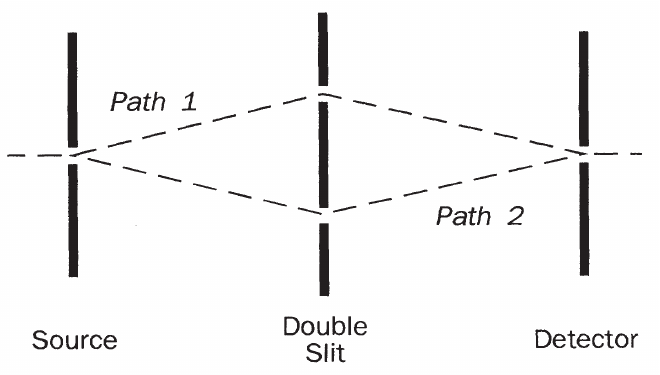}}}\hspace{6pt}
\subfigure[Measured atomic density profile at the detector plane \cite{DSatom}. It was monitored with a 8 $\mu$m grating slit. The dashed line is the detection background.]{
\resizebox*{6cm}{!}{\includegraphics{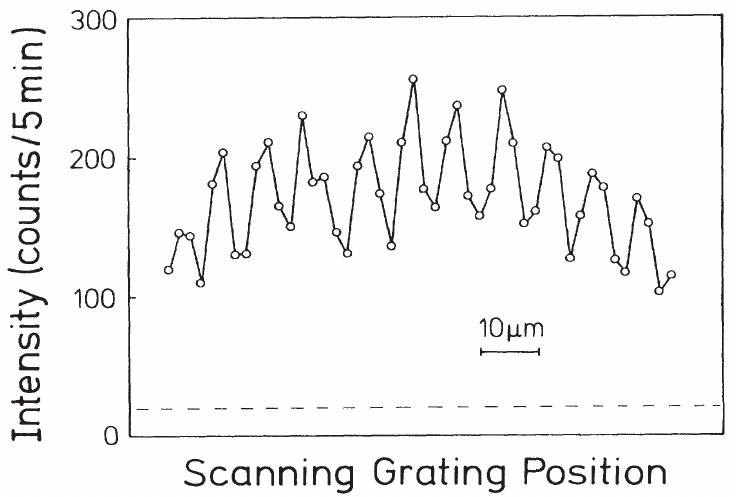}}}
\caption{Young's double-slit experiment with helium atoms. Reprinted  with permission from \cite{DSatom}. Copyright (1991) by the American
Physical Society. } \label{fig:atom}
\end{minipage}
\end{center}
\end{figure}

 Unlike electrons, neutral atoms carry no charge and have shorter wavelengths.
The matter wave interference experiment with atoms was first reported in 1991 \cite{DSatom}. In the experiment, metastable helium atoms were used, which
have a relatively large de Broglie wavelength due to their small mass.  Besides, the metastable helium atoms can be easily detected by an optical method.  The scheme of the experiment is shown
in Figure \ref{fig:atom} (a). An intense atomic beam of helium was produced by supersonic
gas expansion. Most of the atoms ($90\%$) were prepared into  $2 {}^1S_0$ state. The mean velocity $v_0$ of the atomic beam could be controlled by
changing the gas reservoir temperature. Therefore the mean de Broglie wavelength could be controlled. For example, when the reservoir temperature is
$T=83$ K, the de Broglie wavelength is $\lambda_{\mathrm{dB}}=1.03$ \r{A}. The atoms first passed through a slit with a width of $s_1= 2~\mu$m. After
traveling a distance $L=64$ cm, they passed through two $1$-$\mu$m-wide slits, separated by $8~\mu$m. The density profile of atoms was detected
in a plane located another $L'=64$ cm behind the double slit. We expect a modulated intensity distribution with a periodicity $dx= L'\lambda_{\mathrm{dB}}
/d$ and the envelop with a full width of $2L' \lambda_{\mathrm{dB}}/s_2$, where $d$ and $s_2$ are the distance between the slits and the width of the
double slits, respectively. Fig. \ref{fig:atom}(b) shows the experimental results when the wavelength is $\lambda_{\mathrm{dB}}
= 1.03$ \r{A}. The average distance between two maximum is $7.7\pm 0.5~\mu$m, which agrees with the theoretical prediction.

It is more challenging to realize matter wave interferometers for molecules
\cite{DSmole}. The first experimental
paper in this direction was published in 1999 by Zeilinger's group \cite{DSC60}, who realized the double-slit interference experiment using $C_{60}$ molecules.
As the mass of a $C_{60}$ molecule is much larger than that of a helium atom, the de Broglie wavelength of the $C_{60}$ molecular beam is much smaller. It is  only $5$ pm for a $C_{60}$ molecule moves at $100$ m/s .
In this experiment, the slit separation was reduced to $100$ nm, which was much  smaller than the slit separation in the atomic experiment. The interference fringes were
separated around $50~\mu$m at  $1$ m behind the grating.
Later, similar experiments were realized for larger molecules, such as $C_{70}$ \cite{DSC70}, and tetraphenylporphyrin (TPP) \cite{TPP}.
Comparing to atoms, these complex molecules have much larger masses. Thus they can be used to test the quantum superposition
principle and study  decoherence theories in much larger mass regime.

\subsection{Trapped  ions}

While quantum superposition states of atoms and molecules have been demonstrated with Young's double-slit experiments,  a different experimental setup is required to generate  entangled states, which better corresponds to the Schr\"{o}dinger's cat state. The first such experiment was done using trapped ions by C. Monroe {\em et al} about twenty years ago \cite{SchrodingerAtom}.  In that experiment, the internal hyperfine
state of the ion was entangled to the motional state of the ion to create the Schr\"{o}dinger's cat state  \cite{SchrodingerAtom}.

\begin{figure}[btp]
\begin{center}
\includegraphics[width=12cm]{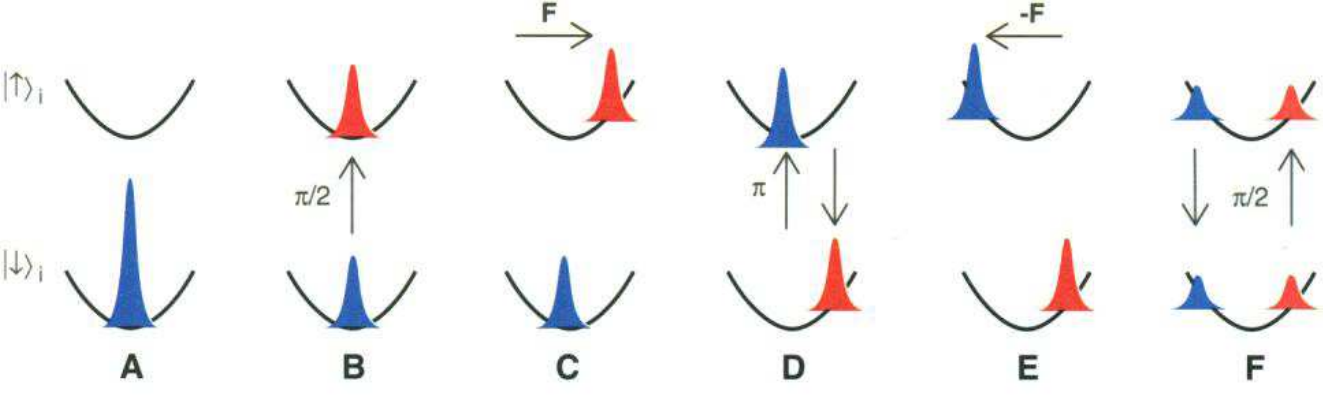}
\caption{Evolution of the spatial atomic wave packet entangled with the internal  hyperfine states $|\downarrow\rangle$ and $|\uparrow
\rangle$. The area of the wave packets corresponds to the probability of  the atom at the given internal state. (A) The ion is initially
at the motional quantum ground state and internal state $|\downarrow\rangle$. (B) The wave function is split by a $\pi/2$ pulse on the internal level.
(C) The $|\uparrow\rangle$ wave packet is excited to a coherent state $|\alpha=3\rangle$ with force \textbf{F}. (D) The $\downarrow\rangle$
and $|\uparrow\rangle$ wave packets are exchanged by a $\pi$-pulse. (E) The $|\uparrow\rangle$ wave packet is driven to a coherent
state $|\alpha=-3\rangle$ with force -\textbf{F}. This state corresponds to a Schr\"{o}dinger's cat state. (F) The $|\downarrow\rangle$
and $|\uparrow\rangle$ wave packets are combined by a $\pi/2$ pulse on the carrier for detection.  Figure adapted from \cite{SchrodingerAtom}. Reprinted with permission from AAAS.
} \label{fig:ioncat}
\end{center}
\end{figure}

As shown in Fig. \ref{fig:ioncat}, a trapped $^9$Be$^+$ ion  was first cooled to the quantum ground state by laser cooling. Then a pair of off-resonant lasers
were used to coherently manipulate its internal   hyperfine  and external motional states. The preparation was divided into $6$ steps.
After the step (E), the atom was prepared into the following entangled state \cite{SchrodingerAtom}:
\begin{equation}
  \label{eq:schrodingeratom}
  \Psi = \frac{|x_1\rangle |\uparrow\rangle + |x_2\rangle |\downarrow\rangle}{\sqrt{2}},
\end{equation}
where coherent states $|x_1\rangle=|\alpha e^{-i\phi/2} \rangle$ and $|x_2\rangle=|\alpha e^{i\phi/2}\rangle$ denote classical-like spatial wave packet states of the trapped ion,
$|\downarrow\rangle$ and $\uparrow\rangle$ represent the internal  hyperfine  states of the trapped ${}^9 Be^+$ ion \cite{SchrodingerAtom}.
The position separation of $|x_1\rangle$ and $|x_2\rangle$ was around $80$ nm, which was larger than both the size of the individual
wave packets ($7$ nm) and the size of the atom ($1$ \r{A}). The mean number of vibrational quanta was $\langle n \rangle = \alpha^2$.
In order to verify the superposition of the Schr\"{o}dinger's cat state,  the coherent wave packets were recombined in the step (F) to
 the following state
\begin{equation}
\Psi= |\downarrow\rangle|S_+\rangle -i |\uparrow\rangle |S_+\rangle,
\end{equation}
with
\begin{equation}
 |S_\pm \rangle = \frac{|\alpha e^{-i\phi/2}\rangle \pm e^{i\delta}|\alpha e^{i\phi/2}\rangle}{2}.
\end{equation}
The populations of $|\downarrow\rangle$
and $|\uparrow\rangle$ depended on the motional phase difference $\phi$ between the two wave packets. Thus the interference between two wave packets could be measured by detecting the probability that the ion was in $|\downarrow\rangle$ (or $|\uparrow\rangle$ ) internal state.

The key features of this experiment \cite{SchrodingerAtom} include: (i) The motion of the trapped ion was controlled with well defined amplitude and
phase;  (ii) the  spatial spreading of individual wave packets
of the atomic motion is small compared to the separation between the two wave packets.  As we will see, the idea of this experiment has been adopted in quantum optomechanics and electromechanics.

\subsection{Quantum optomechanics and electromechanics}
Optomechanics (electromechanics) studies the interaction between light (circuit current) and a mechanical resonator.
Cooling a macroscopic mechanical resonator to the quantum regime, with a mean thermal phonon number less than $1$, is one of the core
issues in quantum optomechanics and electromechanics. The quantum ground state cooling requires the final temperature $T < \hbar \omega_m/k_B$, where $\omega_m$ is the angular frequency of the mechanical oscillator, $\hbar=h/2\pi$, and $k_B$ is the Boltzmann constant.
If  a mechanical resonator has a high resonant frequency, e.g. 1 GHz with $\hbar \omega_m/k_B \sim 50$ mK,  the quantum regime can be approached by
 putting it into a state-of-the-art cryogenic refrigerator ($T\sim 20$ mK). For  mechanical systems with resonant frequencies less than 1 GHz,  other cooling  methods such as cavity sideband cooling
are required to bring them to the quantum regime.

\begin{figure}[btp]
\begin{center}
\begin{minipage}{110mm}
\subfigure[Scanning electron micrograph of a suspended film bulk dilatational resonator at 6 GHz.
]{
\resizebox*{4cm}{!}{\includegraphics{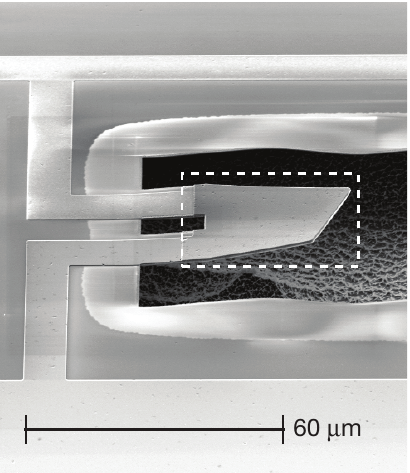}}}\hspace{12pt}
\subfigure[Qubit
excited-state probability as a function of interaction time, showing the quantum state exchange between the superconducting qubit and the mechanical resonator.]{
\resizebox*{6.3cm}{!}{\includegraphics{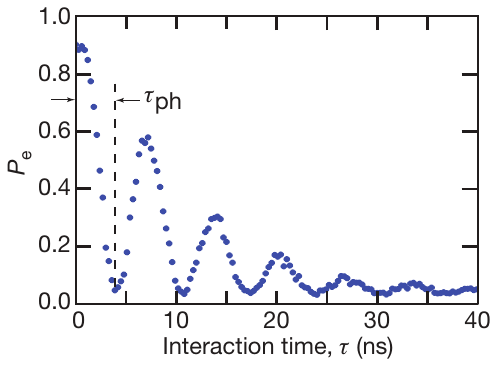}}}
\caption{Ground state cooling and quantum superposition of a mechanical resonator. Reprinted by permission from Macmillan Publishers Ltd: Nature \cite{2010Ground}, copyright (2010).} \label{fig:mech}
\end{minipage}
\end{center}
\end{figure}

In 2010,  quantum ground state cooling of a macroscopic mechanical resonator was realized by conventional cryogenic refrigeration \cite{2010Ground}.
As shown in Figure \ref{fig:mech}a, a micromechanical resonator with a resonant frequency at $6.1$ GHz and a length of about $30~ \mu$m was cooled to $25$ mK with cryogenic refrigeration. The suspended thin film resonator consisted $150$ nm SiO$_2$, $130$ nm Al, $330$ nm AlN and $130$ nm Al.
This $6.1$ GHz mode corresponded to the dilatational vibration (the change of the thickness) of the thin film structure.
As we know, the ground state of this mechanical mode can be reached once its temperature is bellow $0.1$ K.
The AlN film in the structure is piezoelectric. Therefore, the oscillation of the resonator generated a
electric signal, and vice versa.
In order to verify the ground state cooling of the resonator, quantum-limited measurement of the resonator was performed by a superconducting qubit,
which was connected to the resonator through a circuit. The mean thermal phonon number was estimated to be $\langle n_m \rangle <0.07$.
As shown in Figure \ref{fig:mech}b, the qubit was first excited, and then swapped its state with
 the mechanical resonator. The qubit was in its excited state at the maximum point in Fig. \ref{fig:mech}b.  The minimum points corresponded to the situation when the qubit excitation was transferred to the resonator. If the qubit was prepared to be a superposition state initially,  the mechanical resonator would also be in a superposition state by swapping their states. The fitted relaxation time for this mechanical resonator was $6.1$ ns, in good agreement with its measured mechanical quality factor of $Q=260$.

In 2011, sideband cooling of a mechanical resonator to the quantum regime was achieved in both electromechanical and optomechanical systems \cite{2011sideband,2011sideband1}.
In the electromechanical experiment \cite{2011sideband}, a micro-mechanical membrane, with a resonant frequency at $10$ MHz and $Q= 3.3\times 10^5$ was embedded into
a superconducting microwave resonator.  This $10$ MHz mode corresponded to the center-of-mass vibration of the membrane.
The motion
of mechanical resonator would shift the frequency of the microwave resonator.  By  placing the system in a cryogenic refrigerator at $15$ mK and driving the red (low frequency) sideband of the microwave resonator, the mechanical resonator at  $10$ MHz was cooled to the quantum
regime with a mean thermal phonon number $0.34\pm 0.05$. The $100$ nm-thick aluminum membrane vibrated just like the membrane of a drum. Thus we may consider it as a ``quantum drum". Recently, the same group cooled the mechanical mode to a mean thermal
phonon number around $0.1$ and prepared the mechanical mode to  squeezed states by quantum non-demolition measurements \cite{Lecocq2015}. The quantum superposition
of a mechanical resonator has been prepared, and the matter wave interference pattern has been observed in this system.

\section{Experimental developments in quantum teleportation} \label{sec:exptel}

 In scientific fictions, teleportation  describes the hypothetic transfer of an object between two distant locations without
physically moving it along a path. In 1970s, the television series {\em Star Trek} brought the concept of teleportation to  living rooms.
The name of quantum teleportation was inspired by teleportation. However, in quantum teleportation, we can only teleport information,
rather than a physical object \cite{tel}. In other words, quantum teleportation is a form of communication.
In this section, we will introduce the basic quantum teleportation protocol. Then we will discuss  experimental realizations of quantum
teleportation in various systems, such as photons, trapped ions (atoms), and circuit QED.

\subsection{Quantum teleportation protocol} \label{sec:TelPro}
\begin{figure}[btp]
\begin{center}
\includegraphics[width=8cm]{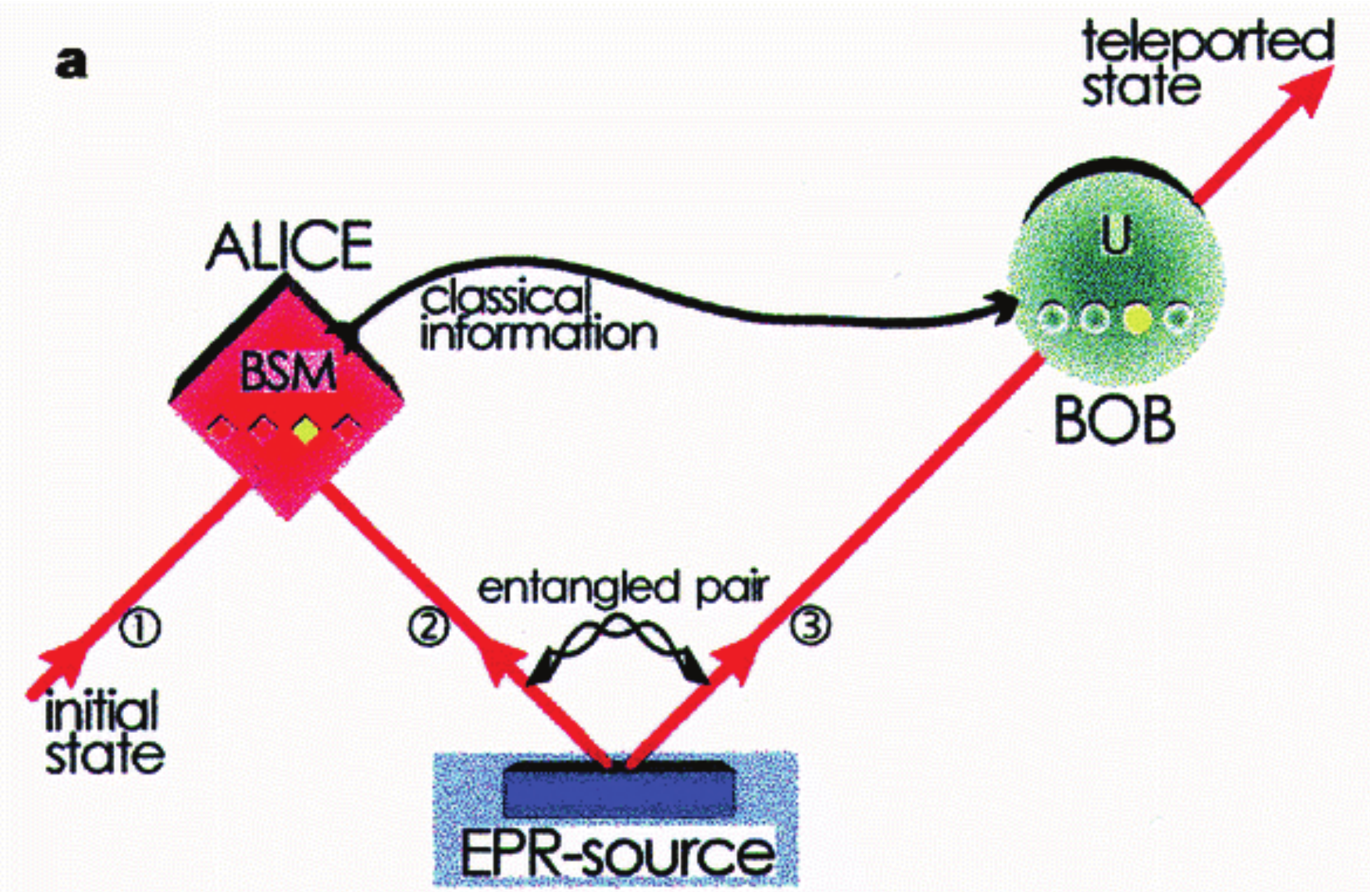}
\caption{Principle of quantum teleportation. Alice has a quantum system (qubit $1$) in an initial state $|\phi_1\rangle$, which she wants to transfer to Bob. An EPR source produces a pair of
entangled photons $2$ and $3$, and sends photon $2$ to Alice, photon $3$ to Bob. Then Alice performs a joint Bell measurement on the initial qubit $1$ and the ancillary photon $2$.
This measurement projects them to one of four different entangled states. She then sends the outcome to Bob though classical channel. Based on the classical information, Bob can perform a local operation on photon $3$ and reproduce the initial state $|\phi_1\rangle$. Bob does not need to know the state $|\phi_1\rangle$ in order to reproduce it.
Reprinted by permission from Macmillan Publishers Ltd: Nature \cite{tel1997}, copyright (1997).}
\label{fig:Tele}
\end{center}
\end{figure}

As we mentioned in Section \ref{sec:concept}, quantum teleportation uses entanglement as a resource to transfer quantum information between
distant locations \cite{tel}. It needs the help of classical communication, but does not need to  move the physical particles in which the quantum information is stored.
After teleportation, the information stored in the original location is destroyed. Therefore, the quantum non-cloning  theorem is obeyed. Here we
discuss the teleportation of a qubit as an example (Fig. \ref{fig:Tele}).

We assume that a qubit state  $|\phi_1\rangle= \alpha |0\rangle_1 + \beta |1\rangle_1$  is stored in  particle 1.
In order to teleport the qubit to another particle, we should prepare an EPR entangled state of a composite system of particles $2$ and $3$. We consider a specific case
\begin{equation}\label{eq:Bell}
  |\Psi_{2,3}^-\rangle = \frac{1}{\sqrt{2}} (|0\rangle_2 |1\rangle_3 - |1\rangle_2 |0\rangle_3).
\end{equation}
Initially, Alice has the qubit $|\phi_1\rangle$. In order to teleport the qubit, one particle (2) of the EPR pair is given to Alice, and the other particle ($3$)
is sent to Bob. Then Alice performs a Bell measurement on both particles $1$ and $2$. The bases of the measurement are
\begin{equation}\label{eq:BellBasis}
  \begin{aligned}
  |\Psi_{12}^\pm \rangle = \frac{1}{\sqrt{2}} (|0\rangle_1 |1\rangle_2 \pm |1\rangle_1 |0\rangle_2), \\
  |\Phi_{12}^\pm \rangle = \frac{1}{\sqrt{2}} (|0\rangle_1 |0\rangle_2 \pm |1\rangle_1 |1\rangle_2).
  \end{aligned}
\end{equation}
The three-particle state before the Bell measurement is
\begin{equation}\label{eq:3state}
  |\Psi_{123}\rangle= \frac{\alpha}{\sqrt{2}} (|0\rangle_1|0\rangle_2 |1\rangle_3 - |0\rangle_1|1\rangle_2 |0\rangle_3)
  +\frac{\beta}{\sqrt{2}} (|1\rangle_1 |0\rangle_2 |1\rangle_3 - |1\rangle_1 |1\rangle_2 |0\rangle_3)
\end{equation}

We can rewrite this three-particle state (Eq. \eqref{eq:3state}) in the Bell bases (Eq. \eqref{eq:BellBasis}):
\begin{equation}\label{eq:BellForm}
  |\Psi_{123}\rangle =\frac{1}{2} \big[ \Psi^-_{12} (-\alpha |0\rangle_3 - \beta |1\rangle_3) +  \Psi^+_{12} (-\alpha|0\rangle_3 + \beta|1\rangle_3)
  +\Phi^-_{12} (\beta|0\rangle_3 + \alpha|1\rangle_3)   +\Phi^+_{12} (- \beta|0\rangle_3+ \alpha|1\rangle_3)     \big]
\end{equation}
Each of the four measurement outcomes in the Bell bases has probability $25\%$. The Bell measurement will project the particle $3$ to one of the four
different states, according to the measurement outcome. The four states are
\begin{equation}
  \label{eq:Bellout}
  |\phi_3^1\rangle = - |\phi_1\rangle, ~ |\phi_3^2\rangle = \left(
                                                             \begin{array}{cc}
                                                               -1 & 0 \\
                                                               0 & 1 \\
                                                             \end{array}
                                                           \right) |\phi_1\rangle,
  ~|\phi_3^3\rangle = \left(
                       \begin{array}{cc}
                         0 & 1 \\
                         1 & 0 \\
                       \end{array}
                     \right) |\phi_1\rangle,
 ~ |\phi_3^4 \rangle= \left(
                       \begin{array}{cc}
                         0 & -1 \\
                         1 & 0 \\
                       \end{array}
                     \right) |\phi_1\rangle.
\end{equation}
These states are simply related to the original qubit state $|\phi_1\rangle$ that Alice has.
Once Alice tells Bob her measurement outcome by classical communication, Bob can apply a local operation on the particle $3$ to restore the qubit $1$.
However, after the Bell measurement, the information in particle $1$ is erased. As the operations in teleportaion are linear, the protocol can be
extended to multiple qubits.

\subsection{Quantum teleportation realizations}

In order to realize quantum teleportation, we need to be able to generate and analyze quantum entanglement, which are challenging.
Several years after quantum teleportation was proposed \cite{tel},  Zeilinger's group realized it with entangled photons \cite{tel1997}.
They generated entangled photon pairs by parametric down-conversion and used two-photon interferometry to analyze entanglement and perform the
Bell measurement.

The principle of quantum teleportaion experiment is shown in Fig. \ref{fig:Tele} \cite{tel1997}. The entangled photons $2$ and $3$ were produced by parametric
down-conversion. An incoming photon went through an nonlinear crystal, which converted it  spontaneously to two photons in the entangled state
$|\Psi_{2,3}^-\rangle= \frac{1}{\sqrt{2}} (|0\rangle_2 |1\rangle_3 - |1\rangle_2 |0\rangle_3)$. Here we denote $|0\rangle$ and $|1\rangle$ as horizontally
and vertically polarized photonic states, respectively.

In order to perform the Bell measurement on photons $1$ and $2$, they must be indistinguishable. Firstly, they were made indistinguishable in time.
Then, the  photons were sent through a spectral filter to reduce their linewidth and enhance their coherence time to be longer than the pump pulse duration.
Finally, they were sent through a 50:50 beam splitter to perform the Bell measurement. The setup in Ref. \cite{tel1997} could only detect  one of the four Bell states. If photons $1$ and $2$ were indistinguishable and in an antisymmetric state, both detectors of the two output ports
of a beam splitter would detect them simultaneously.  In this way, the photons $1$ and $2$ were projected into the antisymmetric state $|\Psi^-_{12}\rangle$, which is the only  Bell state in which the photons go to two ports of the beam splitter.
 Once the Bell measurement was successful, we knew the state of the photon $3$ was $|\phi_3^1\rangle = -|\phi_1\rangle$. So the state of photon 1 was teleported to photon 3.

After the first demonstration, numerous quantum teleportation experiments with photonic systems have been  performed.
The first experiment was done without a complete two-photon Bell measurement. Teleportation could only be realized by selecting the measurement data  after the experiment. In other word, the first teleportation experiment was based on post-selection. In 1998, unconditional quantum teleportation
was proposed and realized  by using squeezed-state entanglement \cite{Braunstein98,Furusawa98}.
Teleporting two degrees  of freedom (spin and orbital angular momentum) of a single photon was demonstrated in 2015 \cite{Wang2015}. The teleportation distance has also been greatly increased
in the last two decades. The longest distance of teleportation is more than $100$ km right now \cite{JYin2014,Ma2014}.

\begin{figure}[tbp]
\begin{center}
\includegraphics[width=15cm]{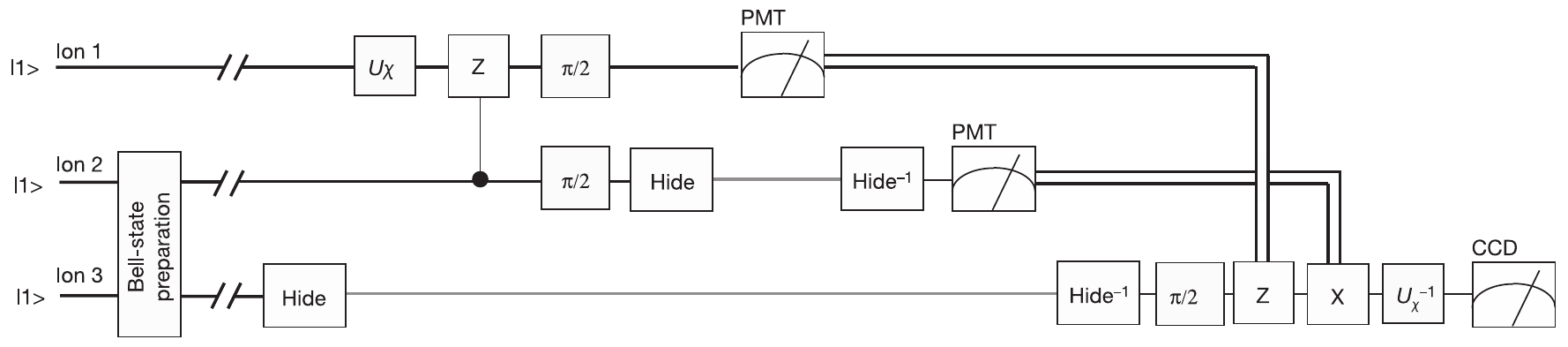}
\caption{Quantum teleportation from ion $1$ to $3$. Ions $2$ and $3$ are initially prepared to a Bell state. The state to be teleported  is encoded in ion $1$ by the operation $U_x$.
In order to realize Bell measurement between ion $1$ and $2$, we need to perform a controlled-Z gate  followed by a $\pi/2$ rotation and
a state detection on ion $1$ and ion $2$. Based on the measurement results (classical information) of both ions $1$ and $2$, the state of ion 1 prepared by the operation $U_x$ could be reconstructed in ion $3$. Double lines in the figure represent the classical information channels. Reprinted by permission from Macmillan Publishers Ltd: Nature \cite{Riebe12004}, copyright (2004). } \label{fig:TeleSolid}
\end{center}
\end{figure}

Comparing to photonic qubits, atomic or solid state qubits can be stored for longer times.
If we can realize quantum teleportation with  atomic or solid state qubits,
the transferred information will be available after the teleportation for further experiments. Besides, the teleportation in these systems could be deterministic, without
post selection. The deterministic
quantum teleportation between ions was reported  by two groups in 2004, by using ${}^9$Be$^+$  \cite{Barrett2004} or ${}^{40}$Ca$^+$ \cite{Riebe12004} ions.
A scheme to teleport the quantum  state of ion $1$ to ion $3$ is shown in Fig. \ref{fig:TeleSolid}.
Unlike photons, we cannot use
 two-particle interferometry to accomplish the Bell measurement because it is difficult to have a beam splitter for atomic and solid state qubits. Fortunately, the Bell measurement can also be realized by performing controlled quantum logic gates
 and measurements on the qubits \cite{Nielsen1998}.

The similar method was later adopted in solid state systems, such as superconducting circuits \cite{Steffen2013}, nitrogen-vacancy (NV) centers in diamond \cite{Pfaff2014}, and etc.
In order to increase the teleportation distance between qubits, photon interference and post-selection methods were used for preparing entanglement between distant
trapped ion or solid state qubits \cite{Olmschenk2009,Pfaff2014}. The entanglement distribution between two NV centers qubits separated by $1.3$ kilometer
has been achieved\cite{Hensen2015}. Loophole-free Bell test has also been performed in this setup \cite{Hensen2015}. With the goal to realize quantum internet in future,
 teleportation between different types of qubits has been studied. Quantum teleportation between light and atomic ensemble \cite{Sherson2006}, between light and solid state quantum memory
 \cite{Gisin2014}, and between photon and phonon \cite{Hou2016} have been demonstrated.

\section{Quantum coherence in biological processes} \label{sec:Qbio}
As discussed in previous sections, quantum coherence and entanglement have been observed in various systems, such as trapped ions, mechanical resonators, superconducting
circuits, and etc. They are crucial resources for quantum information processing. It is natural to ask whether quantum physics plays
a nontrivial role in biology \cite{Lambert2013,Huelga2013}. It is widely known that quantum mechanics is the basic rule of chemical processes. However, it is not clear what is the role of quantum mechanics in biological (physiological) process.  In this section, we will review recent progresses in this direction.
It has been found that  biological systems can perform certain tasks (such as photosynthesis) more efficiently,
or realize a function (such as magnetoreception in some avian species  \cite{Lambert2013}) that cannot be done classically by harnessing  quantum coherence and entanglement.

\subsection{Photosynthesis}

\begin{figure}[bp]
\begin{center}
\includegraphics[width=5cm]{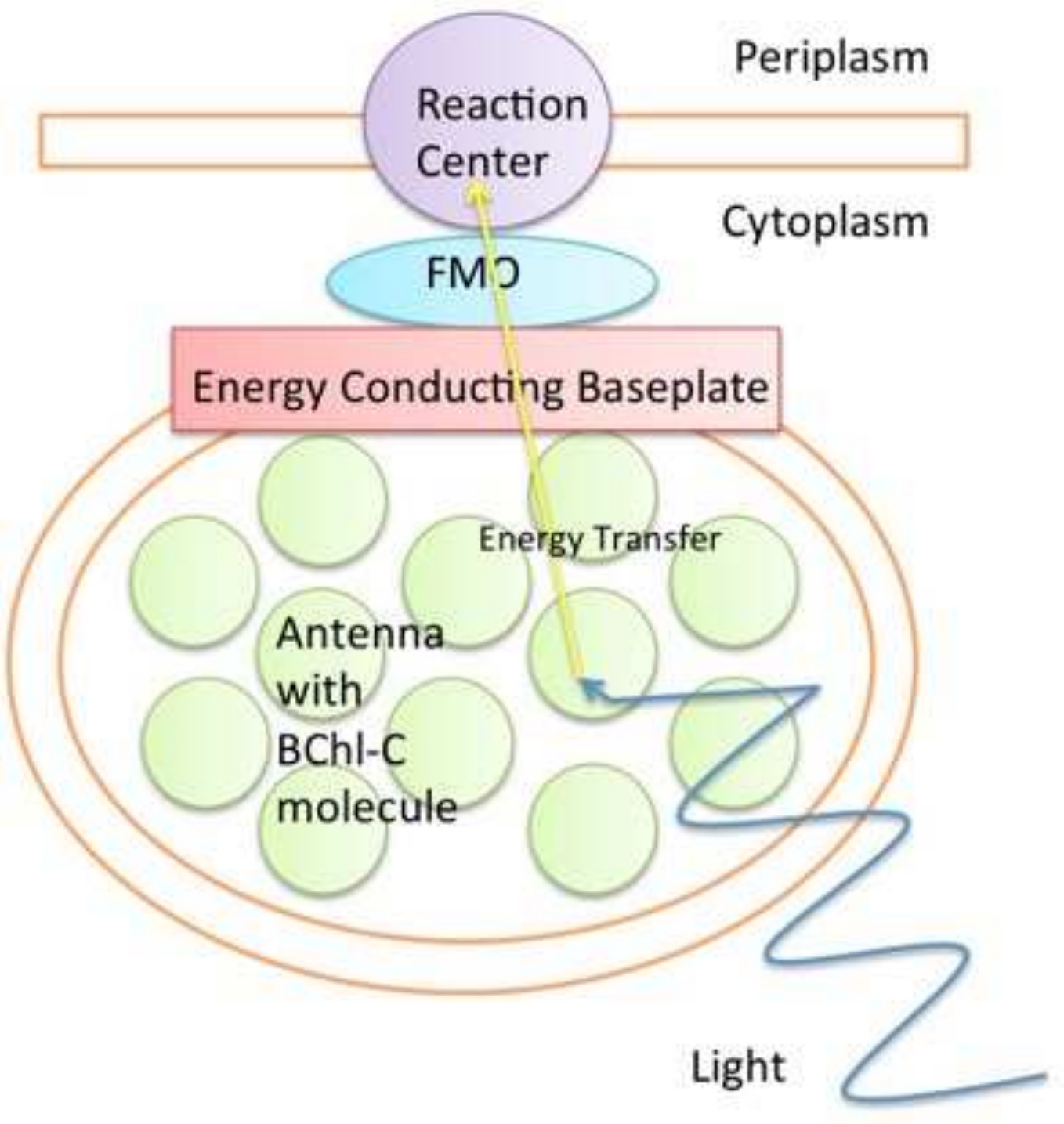} \hspace{10pt}
\includegraphics[width=8.5cm]{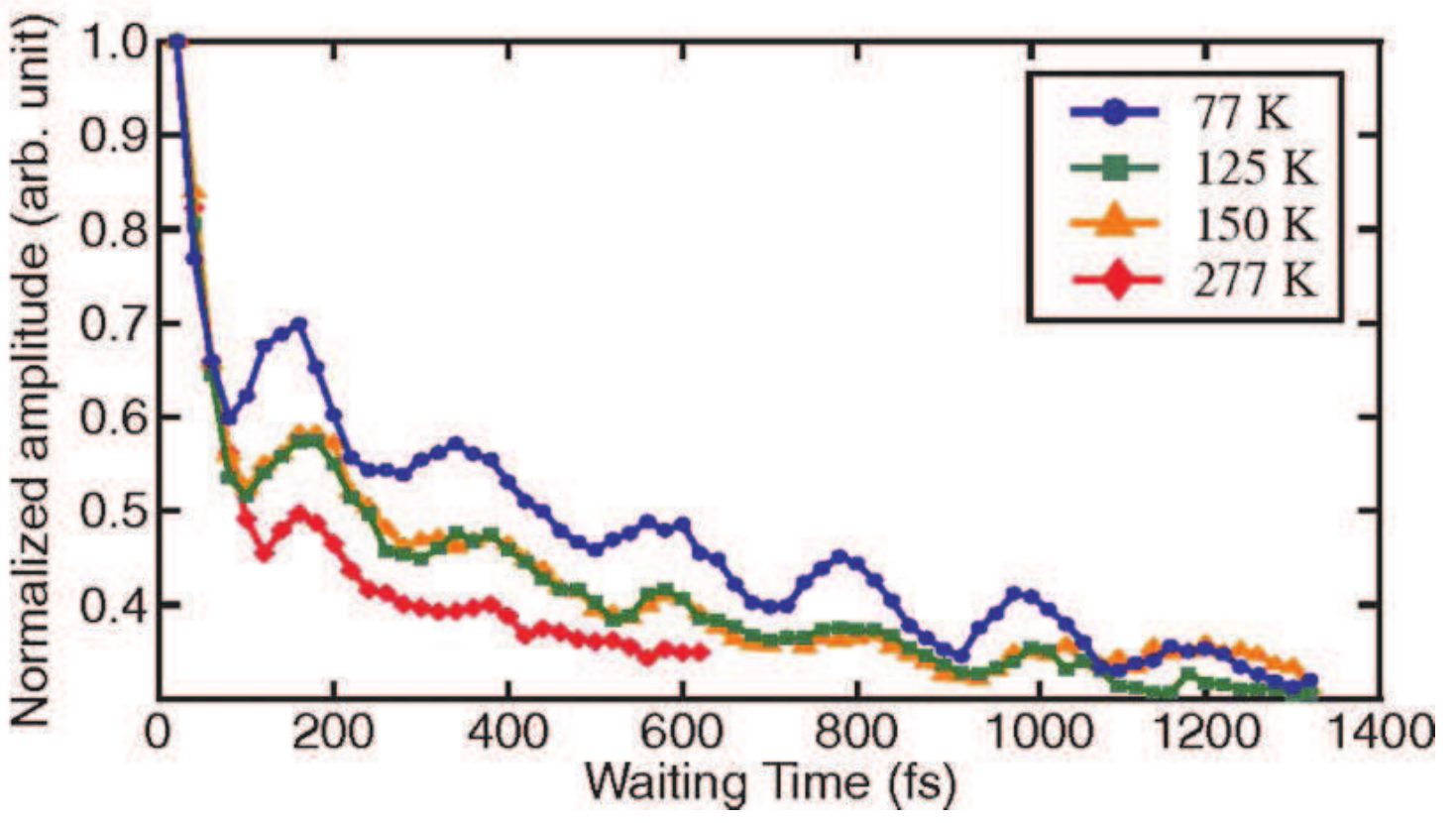}
\caption{The left figure shows  the transfer of an exciton from an antenna that absorbs light to the reaction center in photosynthesis. The right figure shows  that the quantum coherence persists for longer than 300 fs in photosynthesis at 277K. The left figure is adapted from Wikimedia \cite{FMOwiki}, and the right figure is adapted from Ref. \cite{Panit2010}. } \label{fig:photo}
\end{center}
\end{figure}

Photosynthesis is one of the most important biological processes. It provides energy to almost all life on the Earth. In a typical
photosynthetic process as shown in
Fig. \ref{fig:photo}, a photon  is first absorbed by a light-harvesting antenna and creates an exciton.  The exciton is then transferred to a reaction center where the excitation energy is transformed into a more stable chemical energy. The remarkable experimental observation is that almost all of photon energy absorbed by the antenna is transferred to the reaction center. The lifetime of electronic excitation
is only on the order of $1$ ns. Photosynthesis usually performs at temperatures above $273$ K. Therefore, we may treat the exciton transfer as a classical random walk, e.g. the F\"{o}rster model. In this model, the exciton transfer between
different sites is incoherent. The superposition or coherence effects  are  neglected.

In 2007, G. S. Engel {\em et al.} reported that  quantum coherence exists in both the energy transfer in the Fenna-Matthews-Olson (FMO) complex \cite{Engel2007} and the
reaction center in photosynthesis \cite{Lee2007}.
FMO is a specialized structure through which the excitation energy is transferred to the reaction center. The experiment was initially performed
at a cryogenic temperature $77$ K. Later experiments shown that the coherence in excitation transfer exists even near room temperature \cite{Collini2010,Panit2010,Panit2011}, as shown in Fig \ref{fig:photo}. One may ask why quantum coherence exists during the exciton transfer,
and what is the role of it. The simple answer is to increase the energy transfer efficiency. A high transfer
efficiency may be very important for life species that live in a weak light environment, e.g., green sulphur bacteria \cite{Collini2010}.

Numerous theoretical models have been proposed to explain how and why
excitation energy transfer is more efficient by using quantum coherence. For example,
some models treat the environment as a Markovian thermal bath \cite{Mohseni2008,Plenio2009}. Each site in the FMO complex interacts
with an independent environmental bath. It was found that by combining the coherence of excitation transport and the thermal noise, the excitation may
easily escape a local potential minima of the FMO and move to the reaction center.
There were also works on studying the effects of the molecular geometric structure on the efficiency of energy transport  \cite{Ai2012}.
Ref. \cite{Yang2010,Dong2012} studied a dimer structure of light-harvesting complex
2 (LH2), and found that both dimerization and dark states could increase the energy transfer efficiency.

Recent analysis showed that the quantum enhancement of the transport efficiency might be only a few percent \cite{Wu2012}. Thus it is still not clear whether quantum coherence is  essential for
photosynthesis. Some papers proposed classical models that  can also have quantum-like oscillation behaviors \cite{Briggs2011}.
More studies are needed to clarify the role of quantum coherence in photosynthesis \cite{Wilde2010,Tao2016}.

\subsection{Avian magnetoreception}

\begin{figure}[bp]
\begin{center}
\includegraphics[width=10cm]{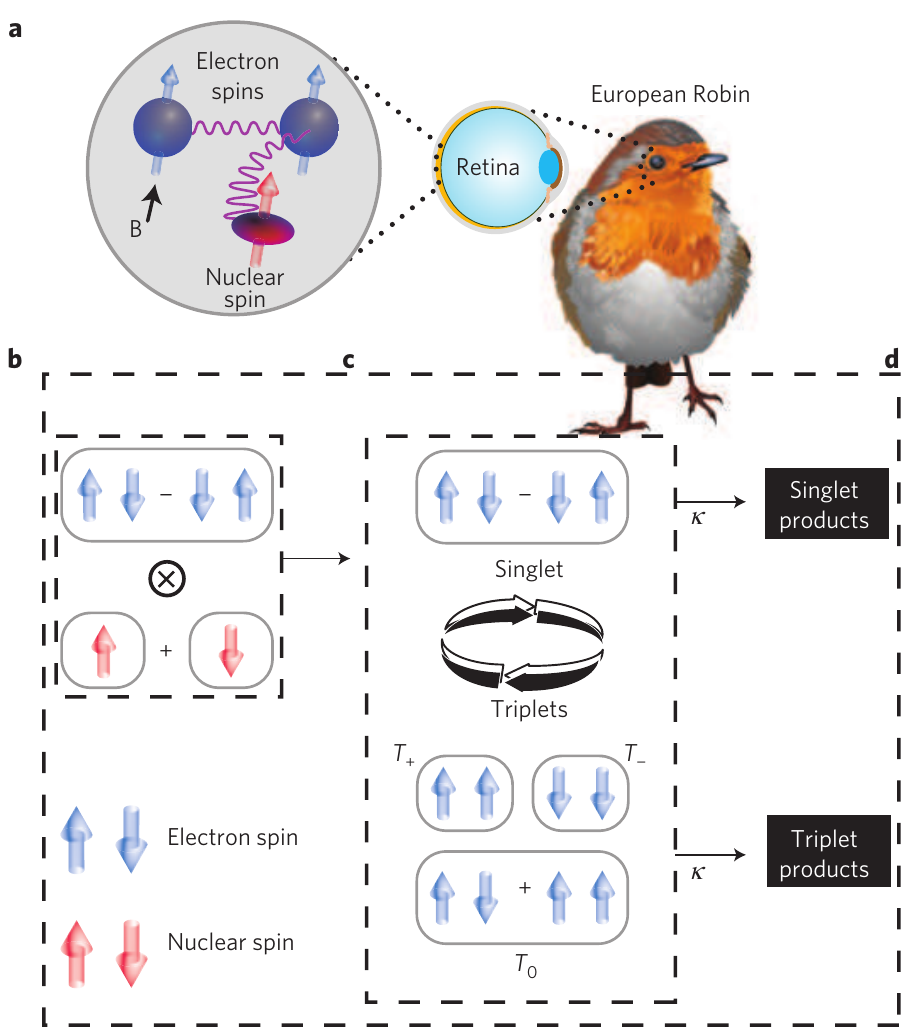}
\caption{The avian quantum compass.(a) A schematic of the RP mechanism for magnetoreception which may explain the navigation
of European robins. RP mechanism is thought to happen in proteins named cryptochromes in the retina. It contains
three main steps. First, light-induced electron transferred from one radial-pair-forming molecule to an acceptor molecule
creates a radical pair. (b,c), Second, the singlet and triplet states convert between each other, which is determined by
the external and internal magnetic couplings. (d) Third, the singlet and triplet pairs recombine into singlet or triplet
products, which could be detected through biological ways. Reprinted by permission from Macmillan Publishers Ltd: Nature Physics \cite{Lambert2013}, copyright (2013). } \label{fig:bird}
\end{center}
\end{figure}

Another widely studied candidate of nontrivial quantum  phenomena in living systems is avian magnetoreception \cite{Wiltschko2006}.
Some migrating animal species  use the weak magnetic field of the Earth for navigation.
The mechanisms of magnetoreception vary from species to species. Behavioral experiments of certain species, such as European robins, show that their navigation is based on both the light  and  the external magnetic field \cite{Ritz2004,Zapka2009}.
These evidences support the so-called radical-pair (RP) mechanism \cite{Lambert2013,Ritz2000}, which is shown in Fig. \ref{fig:bird} \cite{Lambert2013}.

A radical pair is a pair of bound molecules that each has an unpaired electron. As shown in Fig. \ref{fig:bird} \cite{Lambert2013},
the RP mechanism contains three steps. In the first step,  a molecule, such as a cryptochrome protein, in a bird's eye absorbs a photon and generates a
spatially separated electron pair. Usually, the generated pair is in the singlet state before electron transfer.
The singlet  state will evolve due to the interaction with the external magnetic field of the Earth and
the hyperfine interaction with the internal nuclear spins. Then the singlet and triplet states  convert to each other.
Finally, the singlet and triplet pairs recombine. The rate of the recombination
depends on the spin states of the separated charges, which also affect the reaction products of the radical pairs.
The recombination rate should be slower than the singlet-triplet
conversion speed  to allow the RP mechanism to happen. The singlet
and triplet products are  biological detectable in principle.
By detecting the relative weight of the singlet and triplet  products, the angle of the
external magnetic field can be determined. In this way, a magnetic compass is formed in a bird's eye.

Spin singlet and triplet states are highly entangled states, and are  equivalent to Bell states.
There are many studies on how quantum coherence and entanglement will enhance the performance of the RP mechanism.
By using density-matrix equation and quantum measurement theory, I. K. Kominis explained the RP mechanism by
quantum zeno effects \cite{Kominis2009}. J. Cai {\em et al.} studied how quantum control could enhance or
reduce the performance of compass in the RP mechanism, and studied the role of entanglement in this mechanism \cite{Cai2010}.
In the theoretical investigation by E. Gauger {\em et al.} \cite{Gauger2011}, it was found that superposition and entanglement could be
maintained in this system for tens of microseconds in a ``warm and wet'' biological environment. Later, C. Y. Cai {\em et al} found that the sensitivity of the chemical compass based on the RP mechanism
could be greatly enhanced by quantum criticality of the environment \cite{Cai2012} .

There were many experiments that investigated the sensitivity of the RP mechanism in detecting an external magnetic field. In most
experiments \cite{Maeda2008,Lambert2013}, the requirements of high magnetic strength sensitivity (around $50~\mu$T) and high angular sensitivity for avian magnetoreception cannot be fulfilled
at the same time.
Recently, H. G. Hiscock  {\em et al.} studied a modified RP model that involved multinuclear radical pairs \cite{Hiscock2016}.
They found that the output contained a very sharp feature, which could greatly increase the angular sensitivity.
The magnetic strength used in this experiment was also comparable to that of the Earth.


\section{Towards quantum superposition of an optically levitated microorganism} \label{sec:virus}

The quantum coherence in biological processes mentioned in the former section is still at the molecular scale. Can we create quantum superposition of an entire organism as proposed by Schr\"{o}dinger in 1935?  Can we entangle a macroscopic state of an entire living organism to a microscopic  state of an atom or an electron to make a close analog of the Schr\"{o}dinger's cat thought experiment? Remarkably, these long-sought goals in quantum mechanics may be realizable with the state-of-the-art technologies \cite{Svirus,Sbacteria}.

\begin{figure}[bp]
\begin{center}
\includegraphics[width=8cm]{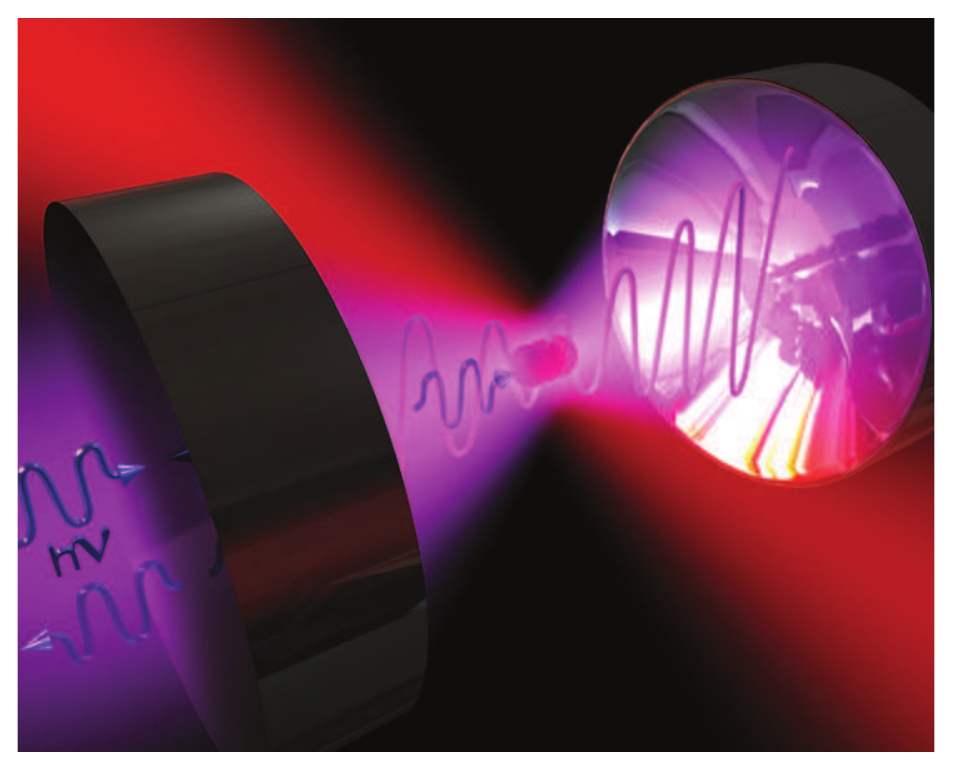}
\caption{An optically levitated microorganism, such as a virus, inside a high-finesse optical cavity in vacuum for creating quantum superposition states. Figure adapted from Ref. \cite{Svirus}} \label{fig:virus}
\end{center}
\end{figure}

In 2009, O. Romero-Isart \emph{et al}. proposed to optically levitate a virus in vacuum inside an optical cavity to create quantum superposition states of a virus (Fig. \ref{fig:virus}) \cite{Svirus}. A virus levitated in high vacuum will be well isolated from the environment. By trapping it in an optical cavity, its motion can be coupled to the photon in the cavity, which can be used to cool its center-of-mass (CoM) motion to the quantum ground state and create superposition states (Fig. \ref{fig:virus}). While Schr\"{o}dinger's cat proposal was a pure thought experiment,
this intriguing proposal of a levitated virus increased our hope to create quantum superposition  of a living organism in a laboratory.
Romero-Isart \emph{et al}. pointed out that this would be possible because (i) living microorganisms have been optically trapped in liquids; (ii) some microorganisms can survive in a vacuum environment; (iii) the size of viruses and some other smallest microorganisms is comparable to the laser wavelength; (iv) some microorganisms are transparent.
Romero-Isart \emph{et al}. proposed that a good example of virus for creating a quantum superposition state will be a tobacco mosaic virus which has a rod-like shape about  50 nm wide and  1 $\mu$m long \cite{Svirus}. Because of its shape, a tobacco mosaic virus will  also be a good candidate to study rotational and torsional cooling.  Currently, the main difficulty to realize this proposal is to optically levitate a virus in high vacuum without significant heating due to light absorption.  The optical absorption coefficients of organisms are typically much larger than that of pure silica optical fibers \cite{Jacques2013}.

While optical levitation of a living microorganism in vacuum has not been realized yet,
there have been many experimental progresses in levitated optomechanics with inorganic dielectric particles \cite{Yin2013a}. In 1975,   A. Ashkin and J. M. Dziedzic optically levitated  glass spheres and oil droplets with diameters about 20-$\mu$m in high vacuum \cite{Ashkin76}. In 2010, Li \emph{et al}. demonstrated feedback cooling of the center-of-mass motion of an optically trapped silica microsphere  from room temperature to about 1.5 mK in high vacuum \cite{Li11}. Parametric cooling \cite{Gieseler12} and cavity cooling \cite{Kiesel13,Asenbaum13,Millen15} of pure dielectric particles (silica and silicon) have also been demonstrated. It is expected that quantum ground state cooing of a levitated pure dielectric particle will be realized soon.

\begin{figure}[btp]
\begin{center}
\includegraphics[width=14cm]{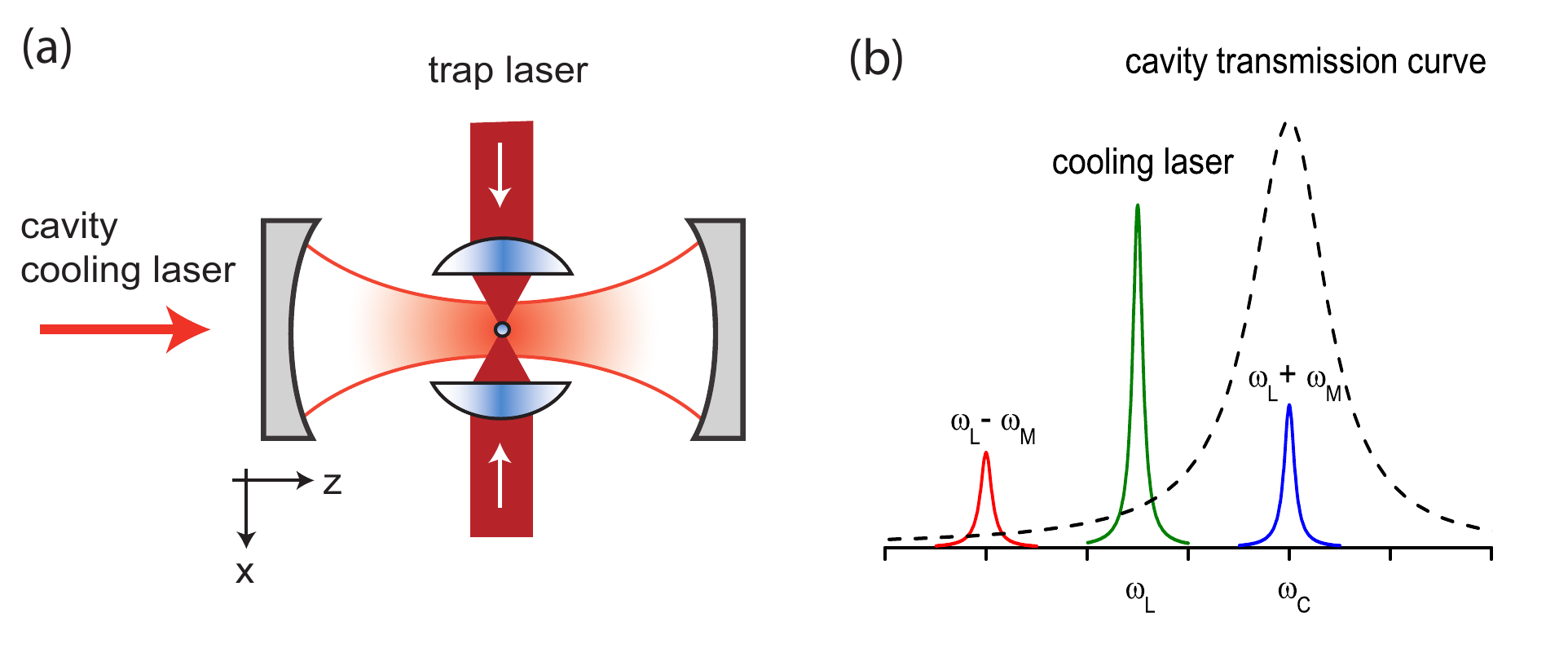}
\caption{(a) Scheme of cavity cooling of a levitated nanoparticle. (b) Principle of cavity cooling. To achieve cooling, the angular frequency of laser $\omega_L$ should be smaller than the resonant angular frequency of the cavity $\omega_C$. $\omega_M$ is the angular frequency of the center-of-mass motion of the levitated nanoparticle. Figure adapted from Ref. \cite{Li13} (With permission of Springer). } \label{fig:cavity1D}
\end{center}
\end{figure}

Let us consider an optically trapped nanoparticle (or a virus) inside an optical cavity as shown in Fig. \ref{fig:cavity1D}(a). The angular frequency of the mechanical vibration of the nanoparticle along the $z$-axis is $\omega_m$. The resonant angular frequency of the cavity without the nanoparticle is $\omega_{C0}$. For a nanoparticle much smaller than the wavelength of the laser, we can use the Rayleigh approximation.
Because of the nanoparticle, the cavity resonant frequency shifts by an amount \cite{Svirus}
\begin{equation}
 \delta \omega_C (\bf r) = -\frac{1}{2} \frac{\int_{V(\bf r)} (\epsilon_r-1)
 {\bf E}^2({\bf r'}) d^3 {\bf r'}}{\int
    {\bf E}^2({\bf r'})d^3{\bf r'}} \cdot \omega_{C0},
\end{equation}
where  ${\bf E}({\bf r'})$ is the electric field of the
cavity mode, $\epsilon_r$ is the relative dielectric constant of the nanoparticle, and $V(\bf r)$ is its occupied space. Because the cavity mode is a standing wave,  $\delta \omega_C (\bf r) $ depends on the location ${\bf r}$ of the nanoparticle in the cavity. The amplitude of $\delta \omega_c (\bf r) $ is maximized when the nanoparticle is at an antinode of the cavity mode, and will be  0 if the nanoparticle is at an node of the cavity mode. Because  the frequency shift  depends on the  position of the nanoparticle, the vibration (phonon) of the nanoparticle is coupled to the photon in the cavity. The typical quantum optomechanical coupling is \cite{Svirus}
\begin{equation}\label{eq:Hom}
  H_{OM} =  \hbar g (a_{\rm m }^\dagger +a_{\rm m }) (a_{\rm c }^\dagger+ a_{\rm c }),
\end{equation}
where $a_{\rm m }^\dagger$ ($a_{\rm m }$) are the creation (annihilation) phonon operators,  $a_{\rm c }^\dagger$ ($a_{\rm c }$) are  operators that create (annihilate) a photon in the cavity, and $g=\sqrt{n_{ph}}g_0$ is the coupling strength. Here $n_{ph}$ is the  number of photons in the cavity and $g_0$ is the  coupling strength between a single photon and a  single phonon.

This optomechanical interaction Hamiltonian (Eq. \ref{eq:Hom}) can be used to cool a levitated nanoparticle. To cool the center-of-mass motion of the levitated nanoparticle, we can choose the laser detuning to be  $\Delta \equiv \omega_L-\omega_C=-\omega_M$, where $\omega_L$ is the frequency of the cooling laser, $\omega_C$ is the resonant frequency of the cavity, and $\omega_m$ is the frequency of the center-of-mass motion of the nanoparticle (Fig. \ref{fig:cavity1D}b). By using rotating wave approximation under the condition $\omega_M >> g$, we get the effective interaction Hamiltonian as \cite{Sbacteria}
\begin{equation}\label{eq:Hinteff}
  H_{\rm eff} =  \hbar g a_{\rm m }^\dagger a_{\rm c }+\hbar g a_{\rm m } a_{\rm c }^\dagger.
\end{equation}

As shown in Fig. \ref{fig:cavity1D}b, the motion of the nanoparticle will modulate the laser field in the cavity and generate two sidebands, one at frequency $\omega_L-\omega_M$ and the other at frequency $\omega_L+\omega_M$ \cite{Li13}. If the laser detuning is $\Delta =-\omega_M$, then the high-frequency sideband $\omega_L+\omega_M=\omega_C$ will be on resonant of the optical cavity and can leak out of the cavity. On the other hand, the low-frequency sideband $\omega_L-\omega_M$ is detuned further away from the resonant frequency of the cavity. On average, the emitted photons will have higher energy than the photons entering to the cavity. Thus the photons will carry away the kinetic energy from the nanoparticle. So the motion of the nanoparticle will be cooled. To achieve ground state cooling, we need to satisfy the resolved sideband limit, i.e. $\omega_M > \kappa >> \gamma_M$. Here $\kappa$ is the line width of the optical cavity, and $\gamma_M$ is the decay rate of the mechanical oscillation of the nanoparticle.

A more interesting task is to create a quantum superposition state of a levitated nanoparticle (or virus). An example superposition state will be $|\Psi>=\frac{1}{\sqrt{2}}(|0>+|1>)$, where $|0>$, $|1>$ are the ground and the first excited state of the center-of-mass vibration of the nanoparticle, respectively. A method is to use a single-photon state. The interaction Hamiltonian (Eq. \ref{eq:Hinteff})  can swap the state of the photon in the cavity to the state of the mechanical motion of the levitated nanoparticle. If the single photon is in a superposition state of entering or not entering to the cavity, its superposition state will be mapped to the mechanical motion of the nanoparticle. Thus the nanoparticle will be prepared in a superposition state. Romero-Isart \emph{et al} later proposed a different method to create superposition states with larger spatial separations by using two optical cavities \cite{Romero11a}. In 2013, Yin \emph{et al} proposed to use electron spin-optomechanical coupling to create large spatial superposition states of a levitated nanodiamond \cite{Yin2013,Yin2015b}. Recently, optical trapping and electron spin control of nanodiamonds in vacuum have been demonstrated \cite{Neukirch2015,hoang2015observation,hoang2016torsional}.

\section{Towards quantum superposition, entanglement, and state teleportation of a microorganism
on an electromechanical oscillator} \label{sec:bact}

\begin{figure}[bp]
\begin{center}
\includegraphics[width=10cm]{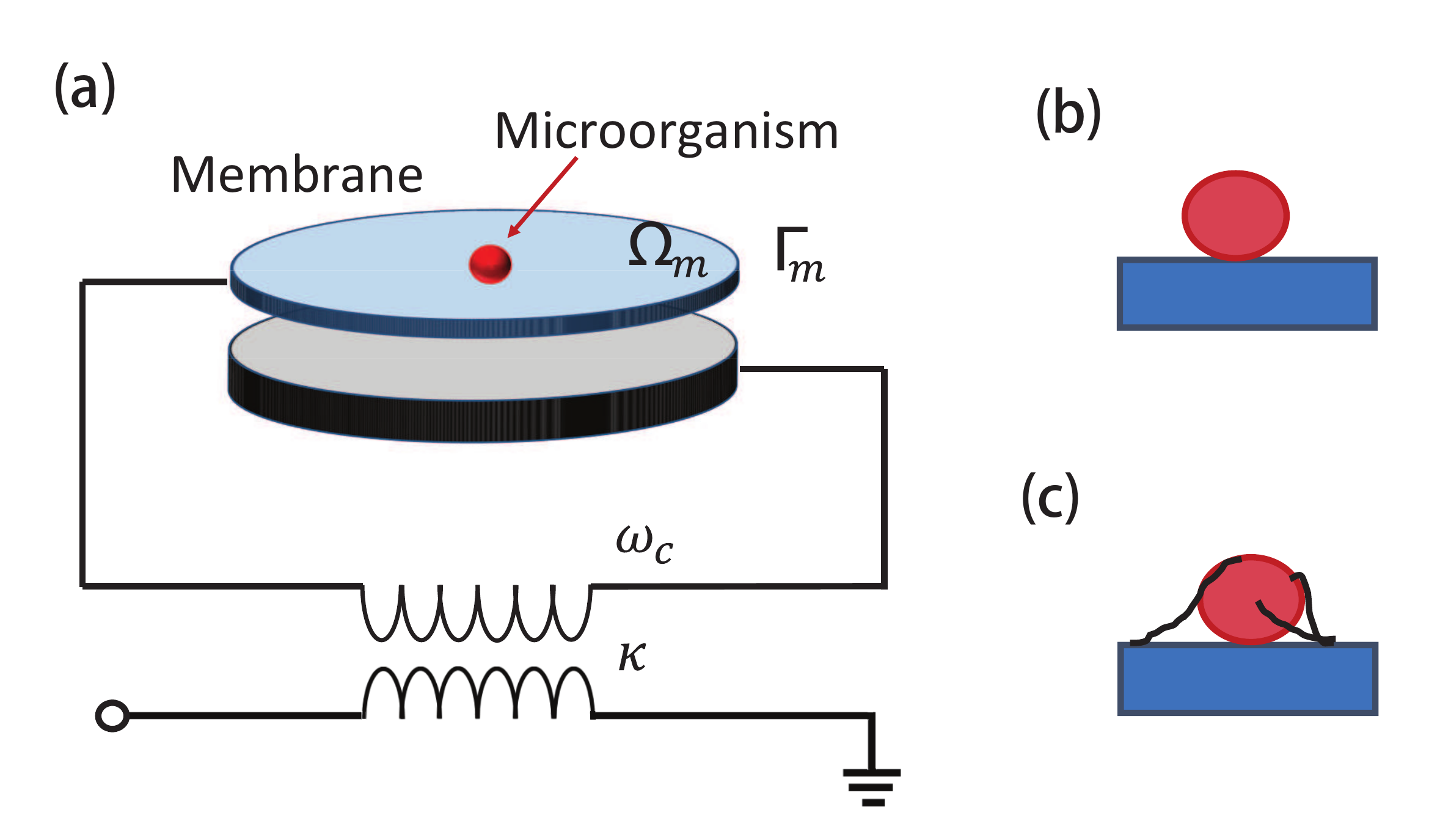}
\caption{Schr\"{o}dinger's microbe. (a) Scheme to create quantum superposition states of a microorganism by putting it on top of an electromechanical membrane coupled to a superconducting LC circuit. (b) A bacterium with a smooth surface. (c) A bacterium with pili on its surface. Figure adapted from Ref. \cite{Sbacteria}.} \label{fig:microbe}
\end{center}
\end{figure}

 In 2015, T. Li and Z.-Q. Yin  proposed to
 create quantum superposition and entangled states  of a  living microorganism by putting a small bacterium  or a virus on top of an electromechanical oscillator, such as a membrane embedded in a superconducting microwave resonant circuit (Fig. \ref{fig:microbe}) \cite{Sbacteria}. By using an electromechanical oscillator instead of optical levitation in vacuum, this approach avoids the heating due to laser absorption.
Electromechanical oscillators imbedded in superconducting circuits have  been cooled to quantum ground state by several groups\cite{2010Ground,2011sideband,Palomaki2013b,Palomaki2013,Suh2012,Suh2014,Wollman2015,Pirkkalainen2015}. Advanced control techniques of superconducting circuits, including quantum teleportation with superconducting circuits, have also been demonstrated \cite{Steffen2013}. In addition, most microorganisms can survive in the cryogenic environment, which is required to use the superconducting circuits.
Microorganisms will be frozen in a cryogenic environment. But they can be still living and  become active after thawing \cite{Mazur1984}.
Cryopreservation is a standard technology for preserving biological samples for long periods and is used clinically worldwide \cite{Mazur1984}. Most microorganisms can be preserved for several years in cryogenic environments without losing vitality \cite{Norman1970}. Even some organs  can  be preserved at cryogenic temperatures \cite{Fahy2004}. At millikelvin temperatures, the sublimation speed of water ice is negligible. It is only  about 0.06 nm  per hour at 128 K \cite{King2005}, and decreases further when the temperature decreases. So a microorganism can be exposed to ultrahigh vacuum without sublimation at millikelvin temperatures.

In the following subsections, we will first review the scheme to create  quantum superposition states of the center-of-mass motion of a microorganism on an electromechanical oscillator. We will then discuss how to create quantum  entanglement between the internal state and the center-of-mass motion of a microorganism. At the end, we will discuss how to teleport the quantum state (center-of-mass motion or internal state) of a microorganism to another microorganism.

\subsection{Center-of-mass motion} \label{sec:CoMbact}

\begin{figure}[btp]
\begin{center}
\includegraphics[width=10cm]{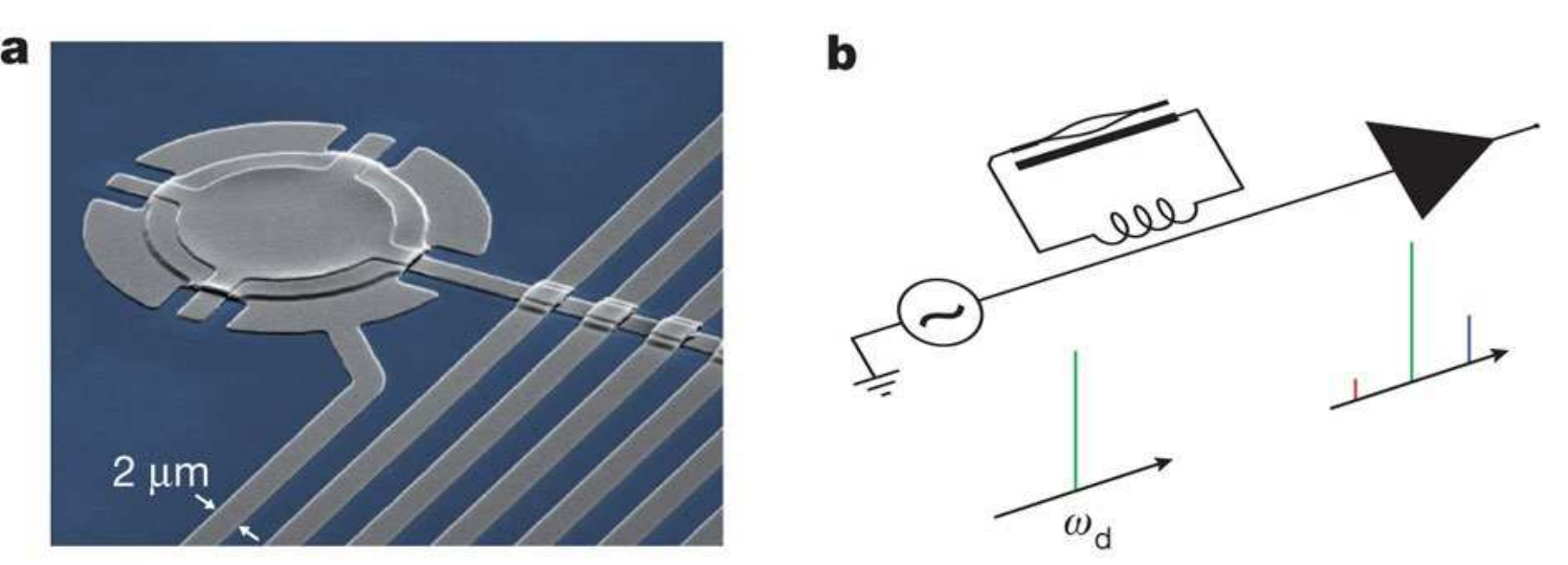}
\caption{An aluminum membrane cooled to the ground state. (a) A scanning electron microscope (SEM) image showing the aluminium (grey) electromechanical circuit. A 15-$\mu$m-diameter aluminium membrane with a thickness of 100nm is suspended 50 nm above a lower electrode. The membrane's vibration  modulates the capacitance of the superconducting microwave circuit. (b) A coherent microwave drive (left, $\omega_d$) inductively coupled to the circuit acquires modulation sidebands (red and blue in the plot below) owing to the thermal motion of the membrane. The upper sideband is amplified with a  Josephson parametric amplifier (filled triangle, right). Reprinted by permission from Macmillan Publishers Ltd: Nature \cite{2011sideband}, copyright (2011).} \label{fig:membrane}
\end{center}
\end{figure}

As proposed in Ref. \cite{Sbacteria}, we can create  quantum superposition states of a small microorganism by putting it on a membrane oscillator (Fig. \ref{fig:microbe}).
Fig. \ref{fig:membrane} shows a 15-$\mu$m-diameter  aluminum membrane with a thickness of 100 nm that has been cooled to quantum ground state by sideband cooling with a superconducting inductor-capacitor (LC) resonator \cite{2011sideband}. The experiment\cite{2011sideband} was performed in a cryostat at 15 mK. The fundamental vibration mode frequency of this membrane is about 10 MHz. Its mechanical quality factor is about $3.3 \times 10^5$. The mass of the membrane is 48 pg ($2.9 \times 10^{13}$ Da), which is larger than those of many microorganisms. Besides ground state cooling, coherent state transfer between the membrane and a traveling microwave field \cite{Palomaki2013}, and entanglement between the motion of the membrane and a microwave field \cite{Palomaki2013b} have been realized. These developments provide the toolbox for creating quantum superposition states of a small microorganism.
 The masses of some common small microorganisms are listed in Table \ref{table1}.  From Table \ref{table1}, it is clear that the mass of the aluminium membrane used in Ref. \cite{2011sideband} is about four orders larger than the mass of ultra-small bacteria \cite{Spirin2002,Fuerstenau2001,Ruigrok1984,Luef2015,Zhao2008,Partensky99,Neidhardt96}.
 A good example of cells that are suitable for performing this experiment is a mycoplasma bacterium. Mycoplasma bacteria are ubiquitous and their sizes are small \cite{Zhao2008}. We can utilize techniques developed in cryo-electron microscopy to prepare the system in a cryogentic environment\cite{Adrian1984}.

As shown in Fig. \ref{fig:microbe}a,  a bacterium or virus can be put on top of an electromechanical membrane oscillator. Its mass $m$ is assumed to be much smaller than the mass of the membrane $M_{\rm mem}$.   For simplicity, it will be better to use microorganisms with smooth surfaces (Fig. \ref{fig:microbe}b), although microorganisms with pili (hairlike structures) on their surfaces (Fig. \ref{fig:microbe}c) will also work as long as the vibration frequencies of the pili are different from the vibration frequency of the superconducting membrane.

\renewcommand{\arraystretch}{1.5}
\begin{table}[bp]
	\centering
\caption{\label{table1} Comparison of  masses of some viruses and bacteria to the mass of a membrane oscillator ($M_{\rm mem}$ = 48 pg) that has been cooled to the quantum ground state in Ref. \cite{2011sideband}.  Table adapted from Ref. \cite{Sbacteria}.  }
		\begin{tabular}{lll}
			\hline
			        {\bf Microorganism}          & {\bf Typical mass}  &  $m / M_{\rm mem}$    \\
                                                 &  (pg)  &   ($M_{\rm mem}$ = 48 pg)  \\  \hline
Bacteriophage MS2        &      $6 \times 10^{-6}$      &       $10^{-7}$        \\
Tobacco mosaic virus &   $7 \times 10^{-5}$     &     $10^{-6}$      \\
 Influenza virus  &   $3 \times 10^{-4}$     &     $10^{-5}$      \\
 WWE3-OP11-OD1 ultra-small bacterium &   0.01    &     $10^{-4}$         \\
     Mycoplasma bacterium&   0.02     &     $10^{-4}$         \\
        Prochlorococcus &   0.3    &     $10^{-2}$         \\
   E. coli bacterium&    1     &   $10^{-2}$      \\ \hline
		\end{tabular}
	
\end{table}

At millikelvin temperatures, the mechanical properties of a frozen microorganism will be similar to a glass particle, while its chemical properties are quite different. A frozen microorganism will stick on the membrane due to attractive van der Waals force.
The pull-off force between a 1 $\mu$m sphere and a flat surface due to van der Waals force is on the order 100~nN \cite{Heim1999}. This is about $10^7$ times larger than the gravitational force on a 1 $\mu$m particle. So the microorganism will move together with the membrane.
The oscillation frequency of the membrane oscillator will change by roughly $-{\mathnormal{\Omega}}_{\rm m } m/(2 M_{\rm mem})$, where ${\mathnormal{\Omega}}_{\rm m }$ is the intrinsic oscillation frequency of the membrane. This small frequency shift  will not significantly affect the ground state cooling of the membrane.
The change of the quality factor $Q$ of the membrane oscillator due to a small microorganism ($m/M_{\rm mem} << 1$) will also be negligible. The frequencies of internal vibration modes of the main body of a small bacterium (larger than 1~GHz for a bacterium smaller than 1~$\mu$m) are much larger than the frequency of the center-of-mass motion of the electromechanical membrane which is about 10 MHz. Thus the internal vibration modes of the main body of a bacterium will not couple  to the center-of-mass vibration of the membrane. If the bacterium has  pili  on its surface (Fig. \ref{fig:microbe}c) \cite{Proft2009}, the situation will be more complex. One can  avoid this problem by embedding the pili in water ice.

We assume the frequency of the center-of-mass motion of  the microorganism and the membrane together to be $\omega_{\rm m }$, which is close to ${\mathnormal{\Omega}}_{\rm m }$.
The motion of the membrane alters the frequency $\omega_c$ of the superconducting LC resonator.
The frequency $\omega_c$ can be approximated with $\omega_c(x) = \omega_0 + G x$, where $x$ is
the displacement of membrane, and $G=\partial \omega_c / \partial x$.
The parametric interaction Hamiltonian  has the form $H_\mathrm{I}=
 \hbar G a^\dagger a \hat{x}= \hbar G\hat{n} x_0 (a_{\rm m } + a_{\rm m }^\dagger)$, where $a$ ($a_{\rm m }$) and $a^\dagger$ ($a_{\rm m }^\dagger$)
are the creation and annihilation operators for LC (mechanical) resonator, $\hat{n}$ is the photon number
operator, and $x_0=\sqrt{\hbar/2M_{\rm mem}\omega_{\rm m }}$ is the zero point fluctuation for mechanical mode.
We denote the single-photon coupling constant $g_0=Gx_0$.
The LC resonator can be  driven strongly with frequency $\omega_{\rm d}$  to enhance the effective coupling between LC resonator and
mechanical oscillator. We assume
that the steady state amplitude $\alpha$ of the LC mode  is much larger than $1$. So the effective coupling strength will increase to $g=\alpha g_0$.
 The detuning can be freely chosen to satisfy the requirements of different applications. For ground state cooling, we choose ${\mathnormal{\Delta }}= -\omega_{\rm m }$ . Then we get the effective interaction Hamiltonian as \cite{Yin2015} $
  H_{\mathrm{eff}} =  \hbar g a^\dagger a_{\rm m } + \hbar g a a_{\rm m }^\dagger
$, which is basically the same as Eq. \ref{eq:Hinteff}.

Once the mechanical mode is cooled down to the quantum regime by cavity sideband cooling \cite{2011sideband}, we can prepare the mechanical superposition state of a bacterium by
the method of quantum state transfer between the mechanical mode and the LC microwave mode.
For example, we can first generate the superposition state $|\phi_0\rangle= (|0\rangle + |1\rangle)/\sqrt{2}$ for LC mode $a$ with assistance
of a superconducting qubit. Here $|0\rangle$ and $|1\rangle$ are vacuum and  $1$ photon Fock state of the LC mode $a$.
After interaction time $t= \pi/g$, the mechanical mode will be in the superposition state $|\phi_0\rangle$.

\subsection{Internal states of a microorganism} \label{sec:Internalbact}

Key features of a microorganism that are different from a glass bead include its ability of metabolism and its complex internal states.  It will be interesting to create superposition state of the internal state of a microorganism \cite{Sbacteria}.  A suitable  internal state of a microorganism is the electron spin of a radical (or transition metal ion) in the microorganism. Radicals are routinely produced during metabolism or by radiation damage. The electron spin of a glycine radical $\textrm{NH}^+_3{\dot{\textrm{C}}}\textrm{HCOO}^-$  has a relaxation time $T_1 = 0.31 $ s and a phase coherent time $T_{\rm M} = 6~ \mu$s at 4.2 K \cite{Hoffmann1995}. The phase coherent time increases when the temperature decreases.  Moreover, we can use universal dynamic decoupling  to increase the coherent time $T_M$ by several orders, approaching $T_1$ \cite{Lange2010}. Thus we  expect the coherent time of the electron spin of a glycine radical to be much longer than 1 ms at millikevin temperatures.
In the original Schr{\"o}dinger's cat thought experiment, the macroscopic ``alive" or ``dead" state of a cat was entangled with the microscopic state of a radioactive atom. As an analog, we can  entangle the  center-of-mass motion of an entire microorganism to a microscopic internal state of the microorganism.

\begin{figure}[btp]
\begin{center}
\includegraphics[width=6cm]{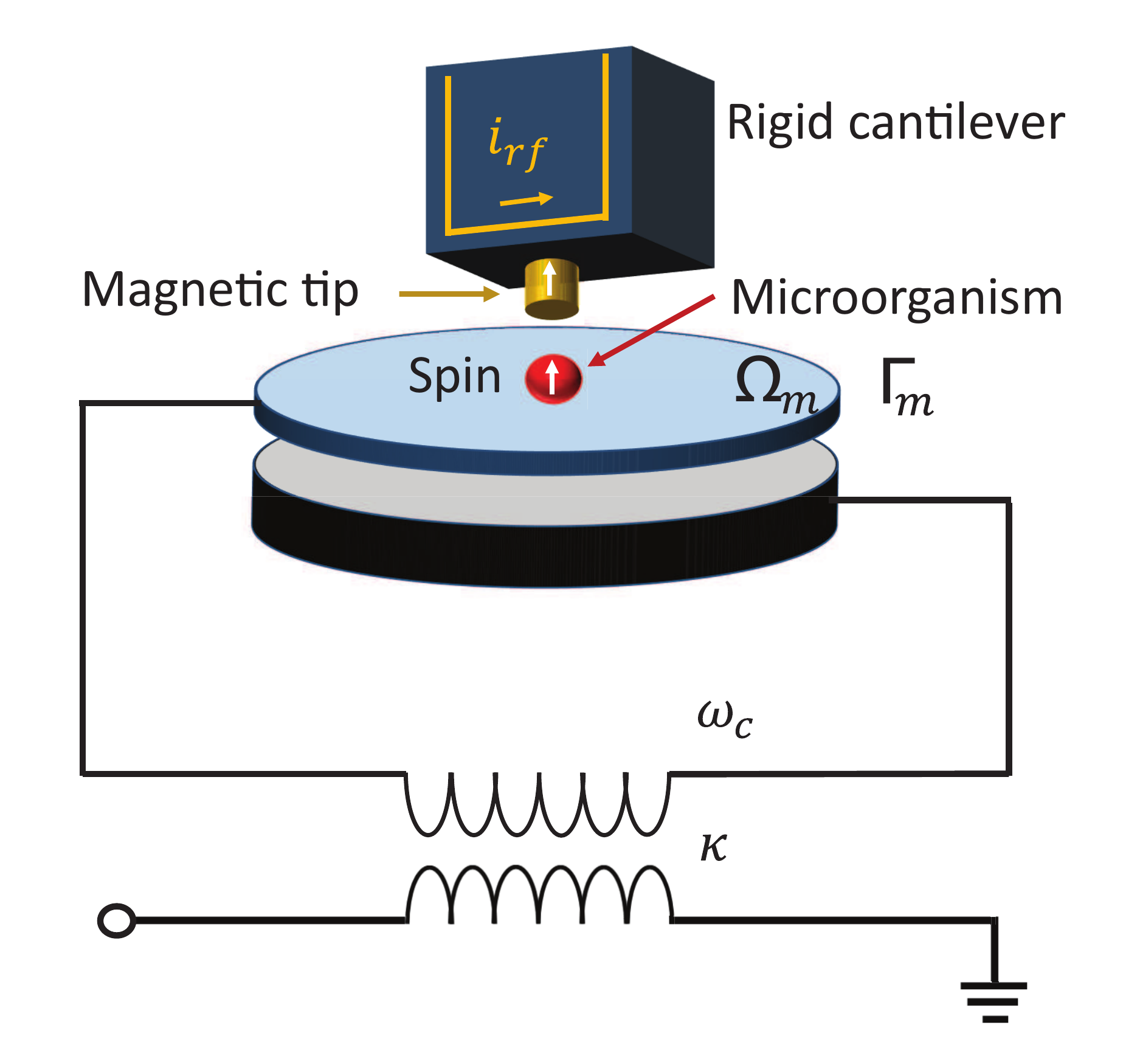}
\caption{Scheme to couple the electron spin of a radial in a microorganism to its center-of-mass motion with a magnetic field gradient created by a nearby ferromagnetic tip. A microwave signal $i_{RF}$ can be used to control the state of the electron spin. The center-of-mass motion of the membrane is also coupled to the microwave photon in the superconducting LC circuit. Figure adapted from Ref. \cite{Sbacteria}.} \label{fig:internal}
\end{center}
\end{figure}

As shown in Fig. \ref{fig:internal}, in order to couple the internal spins states of a  microorganism to the center-of-mass motion of the
microorganism, a magnetic field gradient  is applied. Above the microorganism, there is a ferromagnetic tip mounted on a
rigid cantilever, which produces a magnetic field $\bf{B}$ with a large gradient. This scheme to couple the spin state and the center-of-mass motion of a microorganism  is similar to the scheme used in magnetic resonance force
microscopy (MRFM) \cite{Rugar2004,Degan2009,Vinante2011}. Recently, single electron spin detection with a MRFM  \cite{Rugar2004}, and a MRFM at 30~mK has  been demonstrated \cite{Vinante2011}.

The oscillation of membrane induces a time-varying magnetic field on electrons in the microorganism. We define the
single phonon induced frequency shift $\lambda = g_{\rm s} \mu_{\rm B} |\mathbf{G}_{\rm m }| x_o'/\hbar$, where $x_0'$ is
the zero field fluctuation of microorganism, and $\mathbf{G}_{\rm m }
= \partial \mathbf{B}(\vec{x}_1)/ \partial \vec{x}_1$.  Here we assume that the magnetic gradient is (un)parallel to both the magnetic field $\mathbf{B} (\vec{x}_1)$ and the mechanical oscillation.
The $z$ axis is defined along the direction of $\mathbf{B} (\vec{x}_1)$.
We apply a microwave driving  with frequency $\omega_{\rm d}'$, which is close to the electron $1$'s level spacing $\omega_1=g\mu_{\rm B} B(\vec{x}_1)$.
In a microorganism, there are usually more than one radicals. Because the magnetic field is inhomogeneous, the energy splitting between electron spin states depends on the relative position between an electron and the
ferromagnetic tip.  With the help of a large magnetic gradient, the microwave can be on resonant (or near resonant) of only one electron spin \cite{Rugar2004}. So we can neglect the effects of other electron spins in the microorganism. To make sure the electron spin is initially in the ground state at 10 mK, its energy level spacing should be larger than 500 MHz, which is much larger than the mechanical oscillator frequency.  To achieve strong coupling, we can drive the system with a strong microwave at angular frequency $\omega_1$ with a Rabi frequency $\Omega'_d = \omega_M$. Then the interaction Hamiltonian will be \cite{Sbacteria}
 \begin{equation}\label{eq:HI}
H_{\rm I}= \hbar \lambda \sigma_+ a_{\rm m } + \hbar \lambda \sigma_- a^+_{\rm m },
 \end{equation}
where $\sigma_\pm = \sigma_z \pm {\rm i}\sigma_y$. The spin qubit is defined on the eigenstates of $\sigma_x$.

This interaction Hamiltonian is similar to Eq. \ref{eq:Hinteff}.
 We can generate entangled state and transfer quantum
states between electron spin $1$ and the mechanical mode $a_{\rm m }$ with this Hamiltonian (\ref{eq:HI}).
Similar to the scheme initially proposed for creating quantum superposition state of a nanodiamond with a built-in electron spin, we can prepare the spin of the free electron of a radial to be in a superposition state $|s>=\frac{1}{\sqrt{2}}(|-\frac{1}{2}>+|\frac{1}{2}>)$. With a magnetic field gradient, we can use the interaction Hamiltonian to transfer the spin superposition state to a spatial superposition state, and vice versa. Because of the small vibration amplitude of the high-frequency membrane used in Ref. \cite{2011sideband}, the spatial separation of the two states will be very small ($10^{-14}$ m).
To further increase the spatial separation of the superposition state of a microorganism, one can attach the microorganism to a magnetically levitated superconducting microsphere \cite{Romero-Isart2012b,Cirio2012,Geim1997,Pino2016} instead of a fixed membrane in future. With that method, superposition of two states separated by about 500 nm should be feasible \cite{Pino2016}.

\subsection{Quantum state teleportation of a microorganism} \label{sec:teleportationbact}

\begin{figure}[tbp]
\begin{center}
\includegraphics[width=11cm]{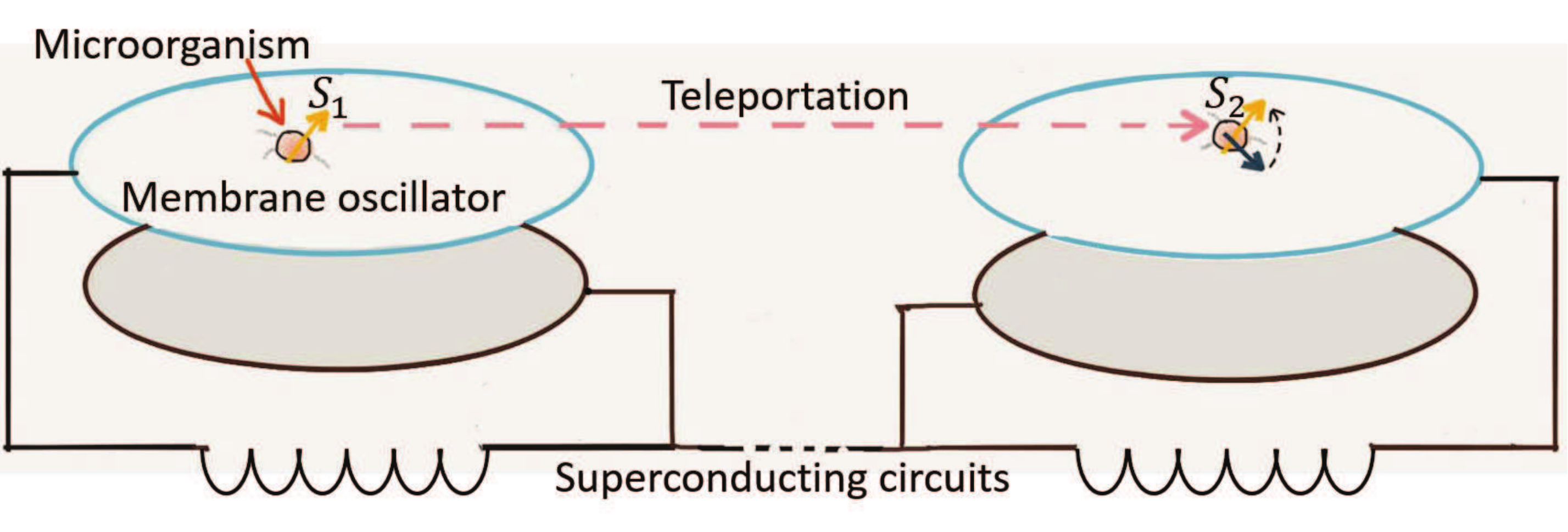}
\caption{Scheme to teleport the quantum state of a microorganism to another microorganism with the help of two electromechanical oscillators and superconducting circuits. Both the center-of-mass motion state and the internal state of the microorganism can be teleported.} \label{fig:teleport}
\end{center}
\end{figure}

 Beyond the Schr{\"o}dinger's cat thought experiment, we can also teleport the center-of-mass motion state and internal electron spin state between two remote microorganisms (Fig. \ref{fig:teleport}) \cite{Sbacteria}. Since internal states of an organism contain  information, this provides an experimental scheme for teleporting  information or memories between two organisms.

 We consider two remote microorganisms, which are
attached to two separate mechanical resonators integrated with LC resonators. They are connected by a superconducting circuit as demonstrated in Ref. \cite{Steffen2013}.  Quantum teleportation based on superconducting circuits has been demonstrated recently \cite{Steffen2013}. The two mechanical resonators will be first cooled down to the motional
ground states. We use the ground state and the first Fock state $|1\rangle$ of both mechanical and LC resonators as the qubit
states. The mechanical mode $a_{m1}$ of the first microorganism and mechanical resonator
is prepared to a superposition state $|\psi_1\rangle = \alpha |0\rangle_{m1}+\beta |1\rangle_{m1}$, where $\alpha$ and $\beta$ are arbitrary and
fulfill $|\alpha|^2 + |\beta|^2=1$. The LC resonator modes $a_1$ and $a_2$ are prepared to the entangled state $(|0\rangle_1 |1\rangle_2+ |1\rangle_1 |0\rangle_2)/\sqrt{2}$, through quantum state transfer, or post selection \cite{Yin2015}. Then
by transferring the state  $a_2$ to the mechanical mode $a_{m2}$ of the second microorganism and mechanical oscillator, the LC mode $a_1$ entangles with mechanical mode $a_{m2}$. This entanglement can be used as a resource for  teleporting the state in mechanical mode $a_{m1}$ to the mechanical mode $a_{m2}$ \cite{tel}. To do this task, we need to  perform  Bell measurements
on mode $a_1$ and $a_{m1}$, which can be accomplished by a CPHASE gate between $a_1$ and $a_{m1}$, and Hadamard gates on $a_1$ and $a_{m1}$ \cite{Steffen2013}.

The internal electron spin state of a microorganism can also be teleported  to another microorganism using superconducting circuits \cite{Steffen2013}. We can first transfer the internal electron spin state of microorganism 1 to its mechanical state with  Hamiltonian (\ref{eq:HI}) \cite{Yin2013,Yin2015b}. We then teleport it to the  mechanical state of the remote microorganism 2. Finally, we transfer the mechanical state of microorganism 2 to its internal electron spin state. In this way, we achieve the quantum teleportation between  internal electron spin states of two  microorganisms. In future,  this method can be extended to entangle and teleport multiple  degrees of freedom \cite{Heilmann2015,Sheng2010,Wang2015} of a living organism at the same time.

\section{Conclusion}

In conclusion, we have reviewed experimental and theoretical progresses related to Schr\"{o}dinger's cat thought experiment. After a brief introduction to basic concepts in quantum mechanics, we review experimental demonstrations of quantum superposition, entanglement and teleportation. We then discuss recent developments in investigating quantum phenomena in living organisms. Several experimental evidences show that quantum coherence and entanglement are important for certain biological functions, such as photosynthesis and magnetoreception. A better understanding of the role of quantum mechanics in photosynthetic processes can also help us to improve the energy efficiency of artificial photosynthetic devices for clean energy. Recently, it was proposed that quantum superposition, entanglement and state teleportation of a microorganism are feasible with state-of-the-art technologies. Once realized experimentally, they would allow us to study some fundamental questions in quantum mechanics, such as the role of observation  and the transition between quantum mechanics and classical mechanics. Quantum teleportation of the states of an organism may find applications in future \cite{teleportorganism}.

%
%
%
%
%

\section*{Acknowledgement}
We thank helpful discussions with Qing Ai. 
Z.Q.Y. is supported by the National Natural Science Foundation of China Grant 61435007. T.L. is supported by the National Science Foundation under Grant No. 1555035-PHY.

\vspace{0.5cm}
\begin{figure}[h]
\includegraphics[height=3cm]{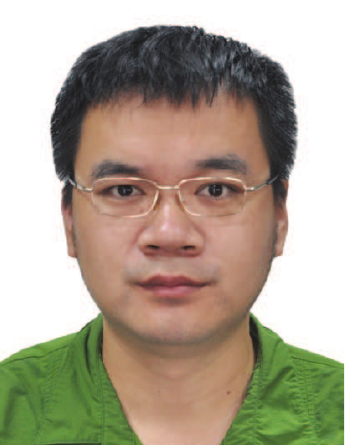}
\caption{{\em Zhang-qi Yin} is an assistant research fellow in the Center for Quantum Information at Tsinghua University in China. He obtained his PhD degree in physics (2009), master degree in theoretical physics (2006), and bachelor degree (2003) from Xi'an Jiaotong University in China. He is interested in quantum information science and the foundation of quantum mechanics, as well as physical implementations of quantum computation and quantum simulation with circuit QED, nitrogen-vacancy centers, and optomechanical systems. } \label{fig:zhangqi}
\end{figure}

\begin{figure}[h]
\includegraphics[height=3cm]{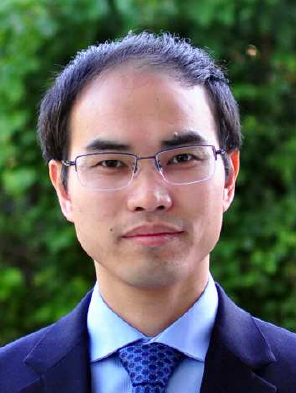}
\caption{{\em Tongcang Li} is an assistant professor of physics and astronomy, and assistant professor of electrical and computer engineering at Purdue University in USA. He obtained his PhD degree in physics from University of Texas at Austin in USA in 2011, and bachelor  degree from University of Science and Technology of China in 2004. He is interested in macroscopic quantum mechanics, quantum optomechanics, quantum optics and plasmonics, laser trapping and cooling, and nonequilibrium thermodynamics. } \label{fig:tongcang}
\end{figure}

\end{document}